\documentclass[longauth,traditabstract]{aa}
\usepackage{epsfig,times}
\usepackage{multirow}
\usepackage{graphicx}
\usepackage{natbib}
\usepackage{txfonts}
\usepackage{lineno}
\usepackage{url}
\usepackage{subfig}
\usepackage{rotating}
\usepackage{tabularx}

\def\mincir{\raise -2.truept\hbox{\rlap{\hbox{$\sim$}}\raise5.truept \hbox{$<$}\ }}
\def\mincireq{\hbox{\raise0.5ex\hbox{$<\lower1.06ex\hbox{$\kern-1.07em{\sim}$}$}}}
\def\magcir{\raise-2.truept\hbox{\rlap{\hbox{$\sim$}}\raise5.truept \hbox{$>$}\ }}
\def\gr{\kern 2pt\hbox{}^\circ{\kern -2pt K}} % ====> GRADI KELVIN

\def\_{\thinspace}

\def\ea{\ et al. \,}

\def\be{\begin{equation}}
\def\ee{\end{equation}}

\newcommand{\Fermic}{\textit{Fermi}}
\newcommand{\Fermi}{\Fermic\ }

\newcommand{\Swiftc}{\textit{Swift}}
\newcommand{\Swift}{\Swiftc\ }

\newcommand{\RXTEc}{\textit{RXTE}}
\newcommand{\RXTE}{\RXTEc\ }

\newcommand{\lapp}{\ensuremath{\stackrel{\scriptstyle <}{{}_{\sim}}}}

\begin{document}

% authors 29.10.2013  Format AA
%
\author{
\small
J.~Aleksi\'c\inst{1} \and
S.~Ansoldi\inst{2} \and
L.~A.~Antonelli\inst{3} \and
P.~Antoranz\inst{4} \and
A.~Babic\inst{5} \and
P.~Bangale\inst{6} \and
U.~Barres de Almeida\inst{6} \and
J.~A.~Barrio\inst{7} \and
J.~Becerra Gonz\'alez\inst{8} \and
W.~Bednarek\inst{9}
% ...
\\
(The MAGIC collaboration) \\ 
B.~Behera\inst{10} \and
M.~Beilicke\inst{32} \and
W.~Benbow\inst{33} \and
K.~Berger\inst{34} \and
R.~Bird\inst{35} \and
A.~Bouvier\inst{36} \and
V.~Bugaev\inst{32} \and
M.~Cerruti\inst{33} \and
X.~Chen\inst{37, 31} \and
L.~Ciupik\inst{38} \and
E.~Collins-Hughes\inst{35} \and
W.~Cui\inst{39} \and
C.~Duke\inst{40} \and
J.~Dumm\inst{41} \and
A.~Falcone\inst{42} \and
S.~Federici\inst{31, 37} \and
Q.~Feng\inst{39} \and
J.~P.~Finley\inst{39} \and
L.~Fortson\inst{41} \and
A.~Furniss\inst{36} \and
N.~Galante\inst{33} \and
G.~H.~Gillanders\inst{43} \and
S.~Griffin\inst{44} \and
S.~T.~Griffiths\inst{45} \and
J.~Grube\inst{38} \and
G.~Gyuk\inst{38} \and
D.~Hanna\inst{44} \and
J.~Holder\inst{34} \and
C.~A.~Johnson\inst{36} \and
P.~Kaaret\inst{45} \and
M.~Kertzman\inst{46} \and
D.~Kieda\inst{47} \and
H.~Krawczynski\inst{32} \and
M.~J.~Lang\inst{43} \and
A.~S~Madhavan\inst{48} \and
G.~Maier\inst{10} \and
P.~Majumdar\inst{49, 50} \and
K.~Meagher\inst{51} \and
P.~Moriarty\inst{52} \and
R.~Mukherjee\inst{53} \and
D.~Nieto\inst{54} \and
A.~O'Faol\'{a}in de Bhr\'{o}ithe\inst{35} \and
R.~A.~Ong\inst{49} \and
A.~N.~Otte\inst{51} \and
A.~Pichel\inst{55} \and
M.~Pohl\inst{37, 31} \and
A.~Popkow\inst{49} \and
H.~Prokoph\inst{10} \and
J.~Quinn\inst{35} \and
J.~Rajotte\inst{44} \and
G.~Ratliff\inst{38} \and
L.~C.~Reyes\inst{56} \and
P.~T.~Reynolds\inst{57} \and
G.~T.~Richards\inst{51} \and
E.~Roache\inst{33} \and
G.~H.~Sembroski\inst{39} \and
K.~Shahinyan\inst{41} \and
F.~Sheidaei\inst{47} \and
A.~W.~Smith\inst{47} \and
D.~Staszak\inst{44} \and
I.~Telezhinsky\inst{37, 31} \and
M.~Theiling\inst{39} \and
J.~Tyler\inst{44} \and
A.~Varlotta\inst{39} \and
S.~Vincent\inst{10} \and
S.~P.~Wakely\inst{58} \and
T.~C.~Weekes\inst{33} \and
R.~Welsing\inst{10} \and
D.~A.~Williams\inst{36} \and
A.~Zajczyk\inst{32} \and
B.~Zitzer\inst{59} \\
(The VERITAS collaboration) \\ 
M.~Villata\inst{60}  \and
C.~M.~Raiteri\inst{60} \and
M.~Ajello\inst{61} \and
M.~Perri\inst{62} \and
H.~D.~Aller\inst{63} \and 
M.~F.~Aller\inst{63} \and
V.~M.~Larionov\inst{64,65,66} \and
N.~V.~Efimova\inst{64,65} \and
T.~S.~Konstantinova\inst{64} \and
E.~N.~Kopatskaya\inst{64} \and
W. P. Chen\inst{67} \and
E. Koptelova\inst{67,68} \and
H. Y. Hsiao\inst{67} \and
O.~M.~Kurtanidze\inst{69,70,79} \and
M.~G.~Nikolashvili\inst{69} \and
G.~N.~Kimeridze\inst{69} \and
B.~Jordan\inst{71} \and
P.~Leto\inst{72} \and
C.~S.~Buemi\inst{72} \and
C.~Trigilio\inst{72} \and
G.~Umana\inst{72} \and
A.~Lahtenmaki\inst{73} \and
E.~Nieppola\inst{73,74} \and
M.~Tornikoski\inst{73} \and
J.~Sainio\inst{20} \and
V.~Kadenius\inst{20} \and
M. Giroletti\inst{75} \and
A.~Cesarini\inst{76} \and
L.~Fuhrmann\inst{77} \and
Yu.~A.~Kovalev\inst{78} \and
Y.~Y.~Kovalev\inst{77,78}
}
\institute { IFAE, Edifici Cn., Campus UAB, E-08193 Bellaterra, Spain
\and Universit\`a di Udine, and INFN Trieste, I-33100 Udine, Italy
\and INAF National Institute for Astrophysics, I-00136 Rome, Italy
\and Universit\`a  di Siena, and INFN Pisa, I-53100 Siena, Italy
\and Croatian MAGIC Consortium, Rudjer Boskovic Institute, University of Rijeka and University of Split, HR-10000 Zagreb, Croatia
\and Max-Planck-Institut f\"ur Physik, D-80805 M\"unchen, Germany
\and Universidad Complutense, E-28040 Madrid, Spain
\and Inst. de Astrof\'isica de Canarias, E-38200 La Laguna, Tenerife, Spain
\and University of Lodz, PL-90236 Lodz, Poland
\and Deutsches Elektronen-Synchrotron (DESY), D-15738 Zeuthen, Germany
\and ETH Zurich, CH-8093 Zurich, Switzerland
\and Universit\"at W\"urzburg, D-97074 W\"urzburg, Germany
\and Centro de Investigaciones Energ\'eticas, Medioambientales y Tecnol\'ogicas, E-28040 Madrid, Spain
\and Technische Universit\"at Dortmund, D-44221 Dortmund, Germany
\and Inst. de Astrof\'isica de Andaluc\'ia (CSIC), E-18080 Granada, Spain
\and Universit\`a di Padova and INFN, I-35131 Padova, Italy
\and Universit\`a dell'Insubria, Como, I-22100 Como, Italy
\and Unitat de F\'isica de les Radiacions, Departament de F\'isica, and CERES-IEEC, Universitat Aut\`onoma de Barcelona, E-08193 Bellaterra, Spain
\and Institut de Ci\`encies de l'Espai (IEEC-CSIC), E-08193 Bellaterra, Spain
\and Finnish MAGIC Consortium, Tuorla Observatory, University of Turku and Department of Physics, University of Oulu, Finland
\and Japanese MAGIC Consortium, Division of Physics and Astronomy, Kyoto University, Japan
\and Inst. for Nucl. Research and Nucl. Energy, BG-1784 Sofia, Bulgaria
\and Universitat de Barcelona (ICC/IEEC), E-08028 Barcelona, Spain
\and Universit\`a di Pisa, and INFN Pisa, I-56126 Pisa, Italy
\and now at Ecole polytechnique f\'ed\'erale de Lausanne (EPFL), Lausanne, Switzerland
\and now at Department of Physics \& Astronomy, UC Riverside, CA 92521, USA
\and now at Finnish Centre for Astronomy with ESO (FINCA), Turku, Finland
\and also at INAF-Trieste
\and also at Instituto de Fisica Teorica, UAM/CSIC, E-28049 Madrid, Spain
\and Now at Stockholms universitet, Oskar Klein Centre for Cosmoparticle Physics
\and now at GRAPPA Institute, University of Amsterdam, 1098XH
Amsterdam, Netherlands % MAGIC
\and Department of Physics, Washington University, St. Louis, MO 63130, USA
\and Fred Lawrence Whipple Observatory, Harvard-Smithsonian Center for Astrophysics, Amado, AZ 85645, USA
\and Department of Physics and Astronomy and the Bartol Research Institute, University of Delaware, Newark, DE 19716, USA
\and School of Physics, University College Dublin, Belfield, Dublin 4, Ireland
\and Santa Cruz Institute for Particle Physics and Department of Physics, University of California, Santa Cruz, CA 95064, USA
\and Institute of Physics and Astronomy, University of Potsdam, 14476 Potsdam-Golm, Germany
\and Astronomy Department, Adler Planetarium and Astronomy Museum, Chicago, IL 60605, USA
\and Department of Physics, Purdue University, West Lafayette, IN 47907, USA 
\and Department of Physics, Grinnell College, Grinnell, IA 50112-1690, USA
\and School of Physics and Astronomy, University of Minnesota, Minneapolis, MN 55455, USA
\and Department of Astronomy and Astrophysics, 525 Davey Lab, Pennsylvania State University, University Park, PA 16802, USA
\and School of Physics, National University of Ireland Galway, University Road, Galway, Ireland
\and Physics Department, McGill University, Montreal, QC H3A 2T8, Canada
\and Department of Physics and Astronomy, University of Iowa, Van Allen Hall, Iowa City, IA 52242, USA
\and Department of Physics and Astronomy, DePauw University, Greencastle, IN 46135-0037, USA
\and Department of Physics and Astronomy, University of Utah, Salt Lake City, UT 84112, USA
\and Department of Physics and Astronomy, Iowa State University, Ames, IA 50011, USA
\and Department of Physics and Astronomy, University of California, Los Angeles, CA 90095, USA
\and Saha Institute of Nuclear Physics, Kolkata 700064, India
\and School of Physics and Center for Relativistic Astrophysics, Georgia Institute of Technology, 837 State Street NW, Atlanta, GA 30332-0430
\and Department of Life and Physical Sciences, Galway-Mayo Institute of Technology, Dublin Road, Galway, Ireland
\and Department of Physics and Astronomy, Barnard College, Columbia University, NY 10027, USA
\and Physics Department, Columbia University, New York, NY 10027, USA
\and Instituto de Astronomia y Fisica del Espacio, Casilla de Correo 67 - Sucursal 28, (C1428ZAA) Ciudad Autónoma de Buenos Aires, Argentina
\and Physics Department, California Polytechnic State University, San Luis Obispo, CA 94307, USA
\and Department of Applied Physics and Instrumentation, Cork Institute of Technology, Bishopstown, Cork, Ireland
\and Enrico Fermi Institute, University of Chicago, Chicago, IL 60637, USA
\and Argonne National Laboratory, 9700 S. Cass Avenue, Argonne, IL 60439, USA 
\and INAF, Osservatorio Astronomico di Torino, I-10025 Pino Torinese (TO), Italy
\and Space Sciences Laboratory, 7 Gauss Way, University of California, Berkeley, CA 94720-7450, USA
\and ASI-Science Data Center, Via del Politecnico, I-00133 Rome, Italy 
\and Department of Astronomy, University of Michigan, Ann Arbor, MI 48109-1042, USA 
\and   Astron.\ Inst., St.-Petersburg State Univ., Russia
\and   Pulkovo Observatory, St.-Petersburg, Russia
\and Isaac Newton Institute of Chile, St.-Petersburg Branch
\and Graduate Institute of Astronomy, National Central University, 300 Jhongda   Rd., Jhongli 32001, Taiwan
\and Moscow M.V. Lomonosov State University, Sternberg Astronomical Institute, Russia
\and Abastumani Observatory, Mt. Kanobili, 0301 Abastumani, Georgia
\and Landessternwarte, Zentrum f\"{u}r Astronomie der Universit\"{a}t
Heidelberg,  K\"{o}nigstuhl 12, 69117 Heidelberg, Germany 
\and School of Cosmic Physics, Dublin Institute For Advanced Studies, Ireland
\and INAF - Osservatorio Astrofisico di Catania, Italy
\and Aalto University Mets\"ahovi Radio Observatory Mets\"ahovintie 114 FIN-02540 Kylm\"al\"a Finland
\and Finnish Centre for Astronomy with ESO (FINCA) University of Turku V\"ais\"al\"antie 20 FIN-21500 Piikki\"o Finland
\and INAF Istituto di Radioastronomia, 40129 Bologna, Italy
\and University of Trento, Department of Physics, I38050 Povo, Trento, Italy
\and Max-Planck-Institut f\"{u}r Radioastronomie, Auf dem H\"{u}gel 69, 53121 Bonn, Germany
\and Astro Space Center of the Lebedev Physical Institute, 117997
Moscow, Russia 
\and Engelhardt Astronomical Observatory, Kazan Federal University, Tatarstan, Russia
\\
{*} Corresponding authors: David Paneque (dpaneque@mppmu.mpg.de),
Konstancja Satalecka (konstancjas@googlemail.com), and Nijil Mankuzhiyil (mankuzhiyil.nijil@gmail.com)
}

\title{Multiwavelength observations of Mrk 501 in 2008}

%\date{Draft version 2013/10/17 }%............ Received ..................; accepted ...................}

 \abstract
   {
{\bf Context:} Blazars are variable sources on various timescales 
over a broad energy range spanning from radio to very high energy
($>100$ GeV, hereafter VHE). Mrk\,501 is one of the brightest blazars at
TeV energies and
has been extensively studied since its first VHE detection in 1996. However,
most of the $\gamma$-ray studies performed on Mrk\,501 during the past
years relate to flaring activity, when the source detection and characterization with the
available $\gamma$-ray instrumentation was easier to perform.

{\bf Aims:} Our goal is to  characterize in detail the
source $\gamma$-ray emission, together with the radio-to-X-ray
emission, during the non-flaring (low) activity,
which is less often studied than the occasional flaring (high) activity.

{\bf Methods:} We organized a multiwavelength (MW) campaign on
Mrk\,501 between March and May 2008. This multi-instrument effort
included  the most sensitive VHE $\gamma$-ray
instruments in the northern hemisphere, namely the imaging atmospheric
Cherenkov telescopes MAGIC and VERITAS, as well as 
\Swiftc, \RXTEc, the F-GAMMA, GASP-WEBT, and other collaborations and
instruments. This provided extensive energy and temporal coverage
of Mrk\,501 throughout the entire campaign.

{\bf Results:} Mrk\,501 was found to be in a low state of activity
during the campaign, with a VHE flux in the range of 10\%--20\% of the Crab nebula
flux. Nevertheless, significant flux variations were detected
with various instruments, with a trend of increasing variability
with energy and a tentative correlation between the X-ray and VHE fluxes.
The broadband spectral energy distribution during the two different
emission states of the campaign can be adequately described within the homogeneous
one-zone synchrotron self-Compton model, with the (slightly) higher
state described by an increase in the electron number density.

{\bf Conclusions:} The one-zone SSC model can adequately describe  the
broadband spectral energy distribution of the source during the two
months covered by the MW campaign. This agrees with 
previous studies of the broadband emission of this source
during flaring and non-flaring states. We report for the first time a
tentative X-ray-to-VHE correlation during such a low VHE activity. 
Although marginally significant, this positive correlation between
X-ray and VHE, which has been reported many times during
flaring activity, suggests that the mechanisms that dominate the
X-ray/VHE emission during non-flaring-activity are not substantially different
from those that are responsible for the emission during flaring activity.
}

   \maketitle
%
%________________________________________________________________

\section{Introduction}
Almost one third of the sources detected at very high energy ($>$100
GeV, hereafter VHE) are BL Lac objects, that is, active galactic
nuclei (AGN) that contain relativistic jets pointing approximately 
in the direction of the observer. Their spectral energy distribution (SED) shows a continuous
emission with two broad peaks: one in the UV-to-soft X-ray band, and 
a second one in the GeV-TeV range. They display no or only very weak
emission lines at optical/UV energies.
One of the most interesting aspects of BL Lacs is their flux
variability, observed in all frequencies and on different timescales ranging from weeks down to minutes, which is often accompanied
by spectral variability. 

Mrk 501 is a well-studied nearby (redshift $z=0.034$) BL Lac that was
first detected at TeV energies by the Whipple collaboration in 1996
\citep{Quinn1996}. In the following years it has been observed 
and detected in VHE $\gamma$-rays by many other Cherenkov telescope
experiments. During 1997 it showed
an exceptionally strong outburst with peak flux levels up to ten times
the Crab nebula flux, and flux-doubling timescales down to 0.5 day
\citep{Aharonian1999}. Mrk 501 also showed strong flaring activity at
X-ray energies during that year. The X-ray spectrum was very
hard ($\alpha<1$, with $F_{\nu} \propto \nu^{- \alpha}$), with the
synchrotron peak found to be at $\sim100~ \mathrm{keV}$, about 2 orders of magnitude higher than in previous
observations \citep{Pian1998}. In the following years, Mrk 501 showed
only low $\gamma$-ray emission (of about 20-30\% of the Crab
nebula flux), apart from a few single flares of higher intensity. In
2005, the MAGIC telescope observed Mrk\,501 during another
high-emission state which, although at a lower flux level than that of
1997, showed flux variations of an order of magnitude and
previously not recorded flux-doubling timescales of only few minutes
\citep{Albert2007}.

  Mrk\,501 has been monitored extensively in X-ray (e.g., Beppo SAX 1996-2001, \citet{Massaro2004}) 
and VHE (e.g., Whipple 1995-1998, \citet{Quinn1999}, and HEGRA 1998-1999,
\citet{AharonianHEGRA2001}), and many studies have been conducted
a posteriori using these observations \citep[e.g.,][]{Gliozzi2006}.  
With the last-generation Cherenkov telescopes (before the new generation of Cherenkov telescopes
started to operate in 2004), coordinated multiwavelength (MW) observations were mostly focused on high VHE activity states
\citep[e.g.,][]{Krawczynski2000, Tavecchio2001}, with few campaigns
also covering  
low VHE states \citep[e.g.,][]{Kataoka1999, Sambruna2000}. The data presented here were taken between March 25 and May 16,
2008 during a MW campaign covering radio (Effelsberg, IRAM,
Medicina, Mets\"ahovi, Noto, RATAN-600, UMRAO, VLBA), optical (through
various observatories within the GASP-WEBT program), UV (\Swiftc/UVOT), X-ray (\RXTEc/PCA, \Swiftc/XRT and \Swiftc/BAT), and
$\gamma$-ray (MAGIC, VERITAS) energies. This MW campaign was the first
to combine such a broad energy and time coverage 
with higher VHE sensitivity and was conducted when Mrk 501 was not in a flaring state.

The paper is organized as follows: In Sect. \ref{sec_instr} we
describe the participating instruments and the data
analyses. Sections \ref{LightCurve}, \ref{VarSec}, and \ref{CorrSec}
are devoted to the multifrequency variability and correlations. 
In Sect. \ref{SEDSec} we report on the modeling of the SED data
within a standard scenario for this source, and in Sect. \ref{discussion}
we discuss the implications of the experimental and modeling results.

\section{Details of the campaign: participating instruments  and temporal coverage}\label{sec_instr}

\begin{figure*}[t]
  \centering
 \includegraphics[width=6.5in]{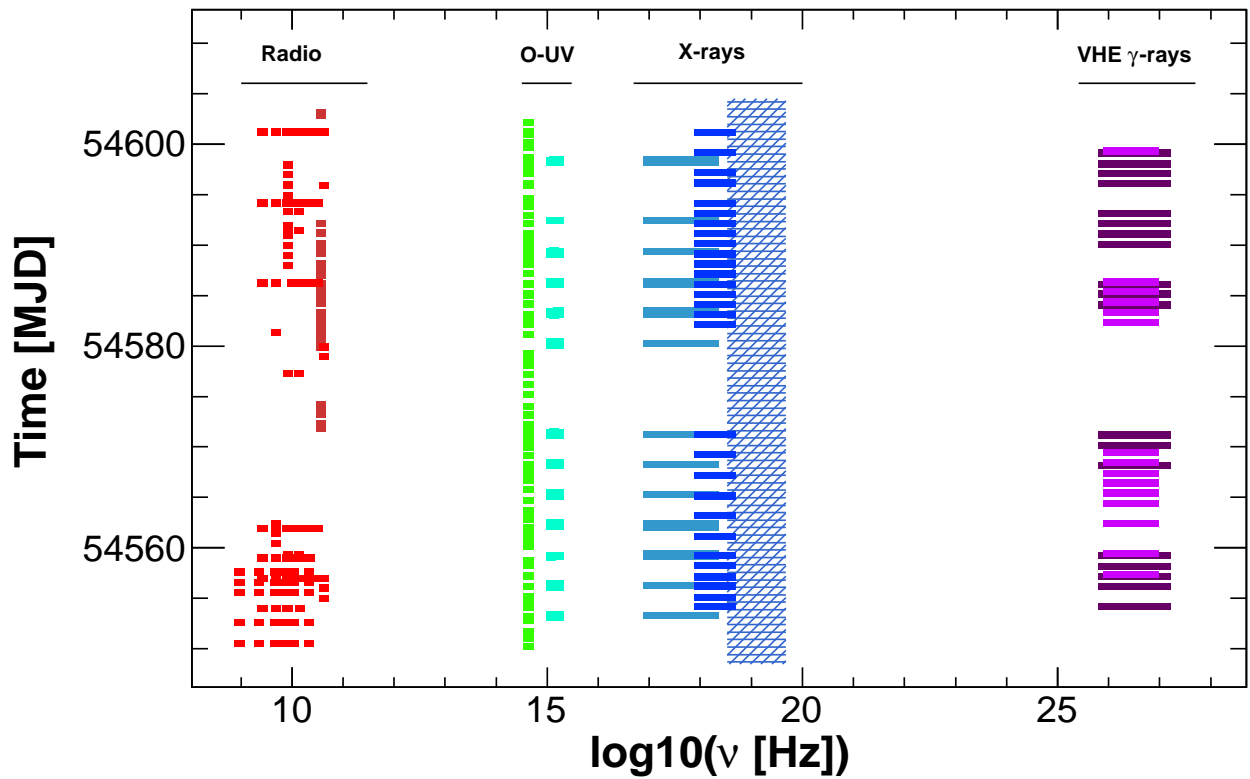}
   \caption{Time and energy coverage during the multifrequency
     campaign. For the sake of clarity, the shortest observing time
     displayed in the plot was set to half a day, and different colors
     were used to display different energy ranges. The correspondence between
     energy ranges and instruments is provided in Table \ref{TableWithInstruments}.
   }
  \label{fig:TimeEnergyCoverage}
\end{figure*}

The list of instruments that participated in the campaign is
reported in Table\,\ref{TableWithInstruments}. Figure\,\ref{fig:TimeEnergyCoverage}
shows the time coverage as a function of the energy range for the
instruments and observations used to produce the light curves presented in
Fig.~\ref{fig:lc} and the SEDs shown in Fig.~\ref{fig:sed}.

%In the following paragraphs we briefly discuss the procedures used in the analysis of 
%the data taken by the different instruments participating in the campaign. 

\subsection{Radio instruments}

In this campaign, the radio frequencies were covered by various 
single-dish telescopes: 
the Effelsberg 100\,m radio telescope, the 32\,m Medicina radio
telescope, the 14\,m Mets\"ahovi radio telescope, the 32\,m Noto radio
telescope,
the 26\,m University of Michigan Radio Astronomy Observatory
(UMRAO), and the 600 meter ring radio telescope RATAN-600.  
Details of the
observing strategy and data reduction are given by \citet[Effelsberg]{Fuhrmann2008,Angelakis2008},
\citet[Mets\"ahovi]{Terasranta1998}, \citet[UMRAO]{Aller1985},
\citet[Medicina and Noto]{Venturi2001}, and \citet[RATAN-600]{Kovalev1999}.

\subsection{Optical instruments}
\label{OpticalInstruments}

The coverage at optical frequencies was provided  by various
telescopes around the world within the GASP-WEBT program \citep[e.g.,][]{Villata2008,
Villata2009}. In particular, the following observatories contributed to this campaign:
Abastumani, Lulin, Roque de los Muchachos (KVA), St. Petersburg,
Talmassons, and the Crimean observatory.
See Table\,\ref{TableWithInstruments} for more details.
All the observations were performed at the $R$ band, using the calibration stars 
reported by \cite{Villata1998}. The Galactic extinction was
corrected for with the coefficients given by \cite{schlegel98}. 
The flux was also corrected for the estimated contribution from the host galaxy, 12 mJy for an aperture radius of 7.5 arcsec \citep{Nilsson2007}.

\subsection{ \Swiftc/UVOT}

The \Swift Ultra Violet and Optical Telescope
\citep[UVOT;][]{Roming2005}
 analysis was performed including all
the available observations between MJD $54553$ and $54599$. The
instrument cycled through each of the three optical pass bands V, B, and U,
and the three ultraviolet pass bands UVW1,
UVM2, and UVW2. The observations were performed with
exposure times ranging from $50$ to $900$ s with a typical exposure of
$150$ s. Data were taken in
the {\it image mode}, where the image is directly accumulated onboard,
discarding the photon timing information, and hence reducing the telemetry
volume.

The photometry was computed using an aperture of $5$ arcsec following
the general prescription of \cite{Poole2008}, introducing an
annulus background region (inner and outer radii 20 and 30 arcsec), and it was corrected for Galactic extinction E(B-V)
= 0.019 mag \citep[]{schlegel98} in each spectral band \citep[]{Fitzpatrick99}.

Note that for each filter the integrated flux
was computed by using the related effective frequency, and not by
folding the filter transmission with the source spectrum. This
might produce a moderate overestimate of the integrated flux of about
10\%. The total systematic uncertainty is estimated to be \lapp18\%.

\subsection{\Swiftc/XRT}

The \Swift X-ray Telescope \citep[XRT;][]{Burrows2005}  pointed
to Mrk 501 18 times in the time interval spaning from MJD
54553 to 54599. Each observation was about 1--2 ks long, with a
total exposure time of 26 ks. The observations were performed in
windowed timing (WT) mode to avoid pile-up, which could
be a problem for the typical count rates from Mrk\,501, which are about
$\sim$5 cps \citep{Stroh2013}.

The XRT data set was first processed with the XRTDAS software package 
(v.2.8.0) developed at the ASI Science Data Center (ASDC) and 
distributed by HEASARC within the HEASoft package (v. 6.13). Event 
files were calibrated and cleaned with standard filtering criteria 
with the {\it xrtpipeline} task. 

The average spectrum was extracted from the summed cleaned event 
file. Events for the spectral analysis were selected within a circle 
of 20 pixel ($\sim$46 arcsec) radius, which encloses about 80\% of the PSF, 
centered on the source position.

The ancillary response files (ARFs) were generated with the 
{\it xrtmkarf} task, applying corrections for the PSF losses and CCD defects 
using the cumulative exposure map. The latest response matrices (v. 014) available 
in the \Swift CALDB\footnote{The CALDB files are located at
  \url{http://heasarc.gsfc.nasa.gov/FTP/caldb}}  
were used. Before the spectral fitting, the 0.3-10 keV source 
energy spectra were binned to ensure a minimum of 20 counts per bin. 
The spectra were corrected for absorption with a neutral hydrogen column density
$N_H$ fixed to the Galactic 21 cm value in the direction of the source, namely 1.56$\times$10$^{20}$cm$^{-2}$ \citep{Kalberla2005}.
When calculating the SED data points, the original spectral data were
binned by combining 40 adjacent bins with the XSPEC command
{\it setplot rebin}. The error associated to each binned SED data
point was calculated adding in quadrature the errors of the original
bins. The X-ray fluxes in the 0.3-10 keV band were retrieved from the
log-parabola function fitted to the spectrum using the XSPEC
command {\it flux}.

\subsection{\RXTEc/PCA}

The {\em Rossi} X-ray Timing Explorer \citep[\RXTEc;][]{RXTERef}
satellite performed 29 pointings on Mrk\,501 during the
time interval  from MJD 54554 to 54601. Each pointing lasted 1.5 ks.
The data analysis was performed using the \texttt{FTOOLS} v6.9 and
following the procedures and filtering criteria recommended by the
\RXTE Guest Observer
Facility\footnote{\url{http://www.universe.nasa.gov/xrays/programs/rxte/pca/doc/bkg/bkg-2007-saa/}}
after September 2007.  The  average net count rate from Mrk\,501 was
about 5\,cps per proportional counter unit (PCU) in the energy range $3-20$\,keV, 
with flux variations typically lower than a factor of two. Consequently,
the observations were filtered
following the conservative procedures for faint sources.
% Earth elevation angle greater 
%than $10^\circ$, pointing offset less than $0.02^{\circ}$, time since the peak of the last 
%SAA (South Atlantic Anomaly) passage greater than 30 minutes, and electron contamination 
%less than $0.1$. 
For details on the analysis of faint sources with \RXTEc, see the 
online Cook Book\footnote{\url{http://heasarc.gsfc.nasa.gov/docs/xte/recipes/cook_book.html}}. 
In the data analysis, only the first xenon 
layer of PCU2 was used to increase the quality of the signal. We used the package \texttt{pcabackest} to model the background, 
the package \texttt{saextrct} to produce spectra for the source and background files 
and the script \texttt{pcarsp} to produce the response matrix.
As with the \Swiftc/XRT analysis, here we also used a
hydrogen-equivalent column density $N_H$ of 1.56$\times$10$^{20}$cm$^{-2}$
\citep{Kalberla2005}. However, since the PCA bandpass starts at 3 keV,
the value used for $N_H$ does not significantly affect our
results. The \RXTEc/PCA X-ray fluxes were retrieved from the
power-law function fitted to the spectrum using the XSPEC
command {\it flux}.

\subsection{\Swiftc/BAT}

The \Swift Burst Alert Telescope \citep[BAT;][]{Barthelmy2005}
analysis results presented in this paper were derived with all the
available data during the time interval from MJD 54548 to 54604.
The seven-day binned fluxes shown in the light curves were determined
from the weighted average of the daily  fluxes reported in the NASA \Swiftc/BAT web page\footnote{
\url{http://swift.gsfc.nasa.gov/docs/swift/results/transients/}}. 
On the other hand, the spectra for the three time intervals defined in
Sect. \ref{LightCurve} were produced following the recipes presented by
 \cite{ajello08,ajello09b}. 
The uncertainty in the \Swiftc/BAT flux/spectra is large
because Mrk\,501 is a relatively faint X-ray source and is therefore
difficult to detect above 15 keV on weekly timescales. 

\subsection{MAGIC}

MAGIC is a system of two 17 m diameter imaging atmospheric Cherenkov
telescopes (IACTs), located at the Observatory
Roque de los Muchachos, in the Canary island of La Palma (28.8 N, 17.8 W, 2200 m a.s.l.).
The system has been operating in stereo mode since 2009
\citep{Aleksic:2011bx}.
The observations reported in this manuscript were performed in 2008,
hence when MAGIC
consisted on a single telescope. 
The MAGIC-I camera contained 577 pixels and had a field of view of 3.5$^{\circ}$. 
The inner part of the camera (radius $\sim$1.1$^{\circ}$) 
was equipped with 397 PMTs with a diameter of 0.1$^{\circ}$ each. 
The outer part of the camera was equipped with 180 PMTs
of 0.2$^{\circ}$ diameter.
MAGIC-I working as a stand-alone instrument was sensitive over an energy range of 50 GeV to 10 TeV 
with an energy resolution of 20\%, an angular PSF of 
about 0.1$^{\circ}$ (depending on the event energy) and a sensitivity
of 2\% the integral flux of the Crab nebula in 50 hr of observation \citep{Albert2008}.

MAGIC observed Mrk\,501 during 20 nights between 2008 March 29 and
2008 May 13 (from MJD 54554 to 54599).
The observations were performed in ON mode, which means that the source
is located exactly at the center in the telescope PMT camera.
The data were analyzed using the standard MAGIC analysis and
reconstruction software MARS \citep{Albert:2007yd, Aliu:2008pd, Zanin2013}.
The data surviving the quality cuts amount to
a total of 30.4 hours. The derived spectrum was unfolded to correct for the
effects of the limited energy resolution of the detector and possible
bias \citep{Albert2007c} using the most recent (March 2014) release of the MAGIC unfolding
routines, which take into account the distribution of the observations in zenith and azimuth
for a correct effective collection area recalculation. 
The resulting spectrum is characterized by a power-law function with spectral index (-2.42$\pm$0.05) 
and normalization factor (at 1 TeV) of (7.4$\pm$0.2)$\times$10$^{-12}$\ cm$^{-2}$s$^{-1}$TeV$^{-1}$ 
(see Appendix \ref{XG_spectra}). The photon fluxes for the individual
observations were computed for a photon index of 2.5, 
yielding an average flux of about 20\% of that of the Crab nebula above 300 GeV, with relatively mild 
(typically lower than factor 2) flux variations.
%For the SED modelling the spectrum was also
%corredted for the extincion on the Extragalactic Background Light (EBL),
%using the emodel of \cite{Franceschini2008}.

\subsection{VERITAS}

VERITAS is an array of four IACTs, each 12 m in diameter, located at
the Fred Lawrence Whipple Observatory in southern Arizona, USA (31.7 N, 110.9 W). Full
four-telescope operations began in 2007. 
All observations presented here were taken with all four telescopes operational, and prior to the relocation of the first telescope within the array layout \citep[]{Perkins2009}. 
%In the summer of 2009 the first telescope was moved to its current
%location in the array to provide a more uniform distance between
%telescopes, improving the sensitivity of the system.
Each VERITAS camera contains 499 pixels (each with an angular diameter
of 0.15$^{\circ}$) and has a field of view of 3.5$^{\circ}$. VERITAS is sensitive over an energy range of 100 GeV to 30 TeV with an energy resolution of 15\%--20\% and an angular resolution (68\% containment) lower than 0.1$^{\circ}$ per event. 

The VERITAS observations of Mrk 501 presented here were taken on 16
nights between 2008 April 1 and 2008 May 13. After applying
quality-selection criteria, the total exposure is 6.2 hr live
time. Data-quality selection requires clear atmospheric conditions,
based on infrared sky temperature measurements, and normal hardware
operation. All data were taken during moon-less periods in wobble mode
with pointings of 0.5$^{\circ}$ from the blazar alternating from
north, south, east, and west to enable simultaneous
background estimation and reduce systematics
\citep[]{AharonianVERITAS2001}. Data reduction followed the methods
described by \cite{AcciariVER2008}. The spectrum obtained with
  the full dataset  is described by a power-law function with spectral index (-2.47$\pm$0.10) 
and normalization factor (at 1 TeV) of (9.4$\pm$0.6)$\times$10$^{-12}$\ cm$^{-2}$s$^{-1}$TeV$^{-1}$ 
(see Appendix \ref{XG_spectra}). In the calculation of the photon
fluxes integrated above 300 GeV for the single VERITAS observations, we used a photon index of 2.5.

\begin{table*}[t]

%  \centering
\begin{tabular*}{\textwidth}{lll}

%\begin{sidewaystable}
%\begin{table}

%\begin{tabular*}{0.9\textwidth}{lll}
\hline 
\hline
Instrument/Observatory & Energy range covered & Web page \\
\hline
MAGIC                 & 0.31-7.0\,TeV               & {\small \url{http://wwwmagic.mppmu.mpg.de/}} \\
VERITAS                 & 0.32-4.0\,TeV               & {\small \url{http://veritas.sao.arizona.edu/} } \\
\Swiftc/BAT                & 14-195\,keV               & {\small \url{http://heasarc.gsfc.nasa.gov/docs/swift/swiftsc.html/} } \\
\RXTEc/PCA                & 3-20\,keV               & {\small \url{http://heasarc.gsfc.nasa.gov/docs/xte/rxte.html} } \\
\Swiftc/XRT                & 0.3-10\,keV               & {\small \url{http://heasarc.gsfc.nasa.gov/docs/swift/swiftsc.html} } \\
\Swiftc/UVOT                & V, B, U, UVW1, UVM2, UVW2            & {\small \url{http://heasarc.gsfc.nasa.gov/docs/swift/swiftsc.html} } \\
Abastumani$^{*}$              & R       band        & {\small \url{http://www.oato.inaf.it/blazars/webt/} } \\
Crimean$^{*}$              & R       band        & {\small \url{http://www.oato.inaf.it/blazars/webt/} } \\
Lulin$^{*}$              & R       band        & {\small \url{http://www.oato.inaf.it/blazars/webt/} } \\
Roque de los Muchachos (KVA)$^{*}$               & R       band        & {\small \url{http://www.oato.inaf.it/blazars/webt/} } \\
St. Petersburg$^{*}$              & R       band        & {\small \url{http://www.oato.inaf.it/blazars/webt/} } \\
Talmassons$^{*}$              & R       band        & {\small \url{http://www.oato.inaf.it/blazars/webt/} } \\
Noto & 43\,GHz & {\small \url{http://www.noto.ira.inaf.it/} } \\
Mets\"ahovi    $^{*}$                   & 37\,GHz               & {\small \url{http://www.metsahovi.fi/} } \\
Medicina & 8.4 \,GHz & {\small \url{http://www.med.ira.inaf.it/index_EN.htm} } \\
UMRAO$^{*}$                & 4.8, 8.0, 14.5\,GHz               & {\small \url{http://www.oato.inaf.it/blazars/webt/} } \\
RATAN-600  &  2.3, 4.8, 7.7, 11.1, 22.2 GHz & {\small \url{http://www.sao.ru/ratan/} } \\
Effelsberg$^{*}$                 &  2.6, 4.6, 7.8, 10.3, 13.6, 21.7, 31\,GHz             & {\small \url{http://www.mpifr-bonn.mpg.de/div/effelsberg/index_e.html/} } \\
\hline
 \end{tabular*}
\caption{List of instruments participating in the multifrequency
  campaign and used in the compilation of the light curves and SEDs
  shown in Fig.~\ref{fig:lc} and \ref{fig:sed}. The instruments
  with the symbol ``$^{*}$'' observed Mrk\,501 through the  GASP-WEBT
  program. The energy range shown in column 2 is 
the actual energy range covered during the Mrk\,501 observations, 
and not the nominal energy range of the instrument, which might only be achievable for bright 
sources and excellent observing conditions.
See text for further comments. }
\label{TableWithInstruments} 
\end{table*}

\section{Light curves}
\label{LightCurve}

\begin{figure*}[H]
  \centering
%\begin{array}{cc}
\includegraphics[width=7.5in]{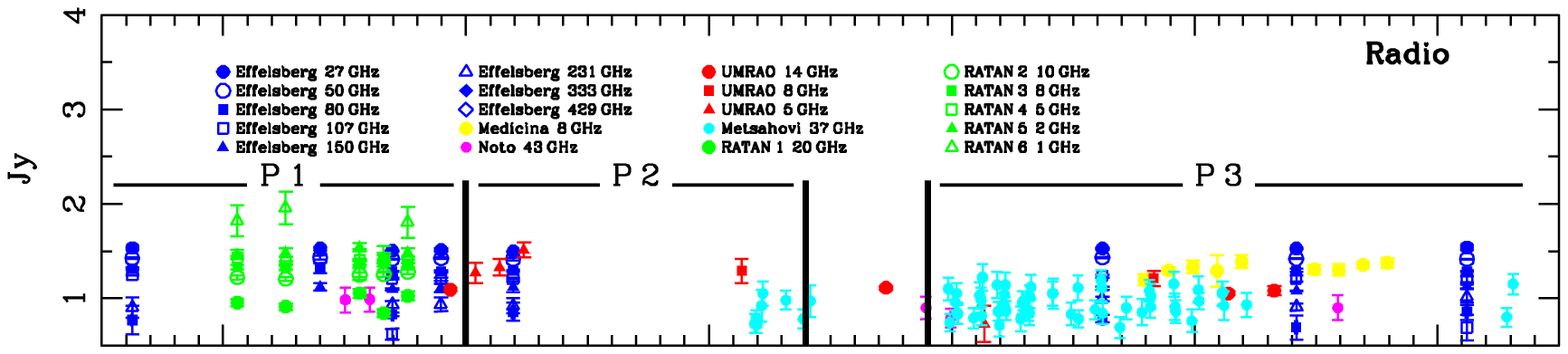} \\
\vspace{-15cm}
\includegraphics[width=7.5in]{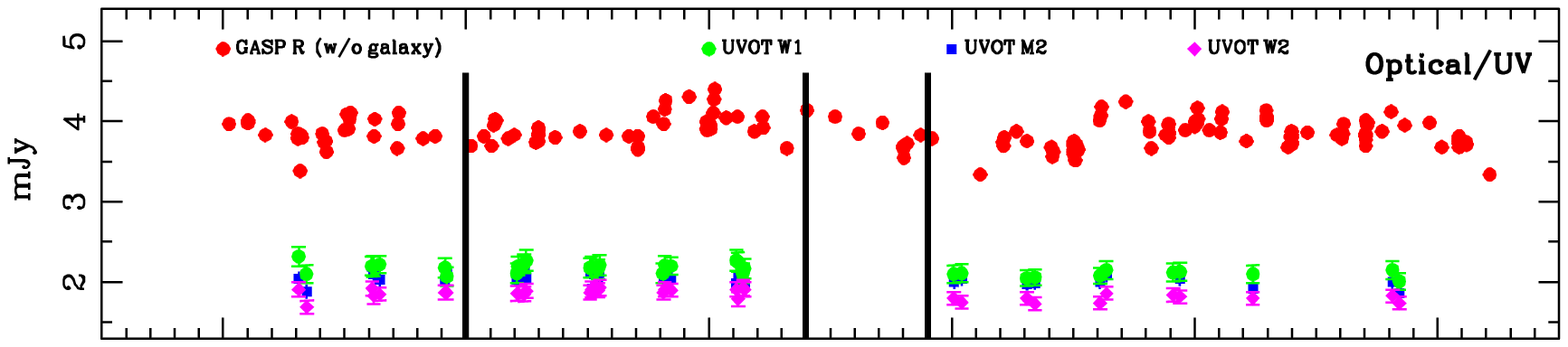}\\
\vspace{-15cm}
\includegraphics[width=7.5in]{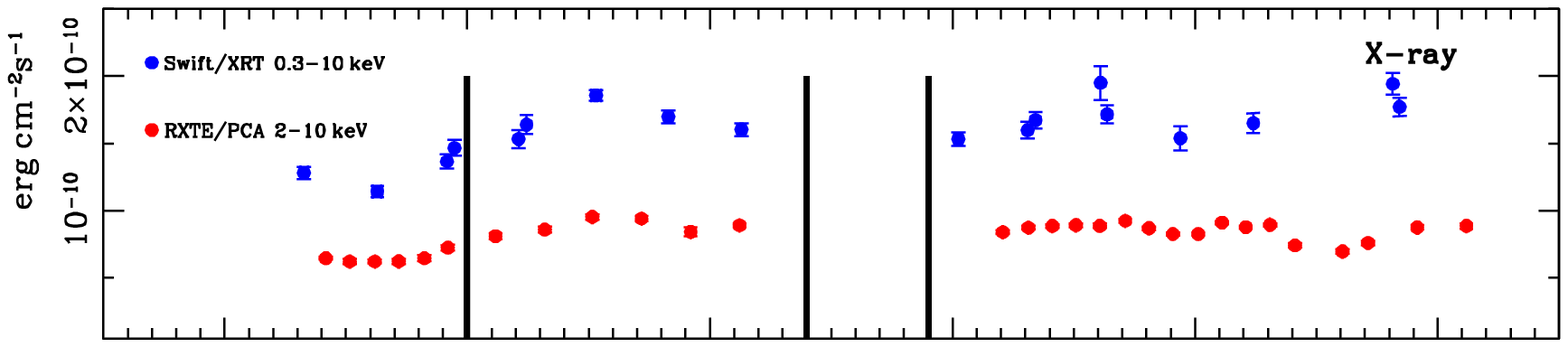} \\
\vspace{-15cm}
\includegraphics[width=7.5in]{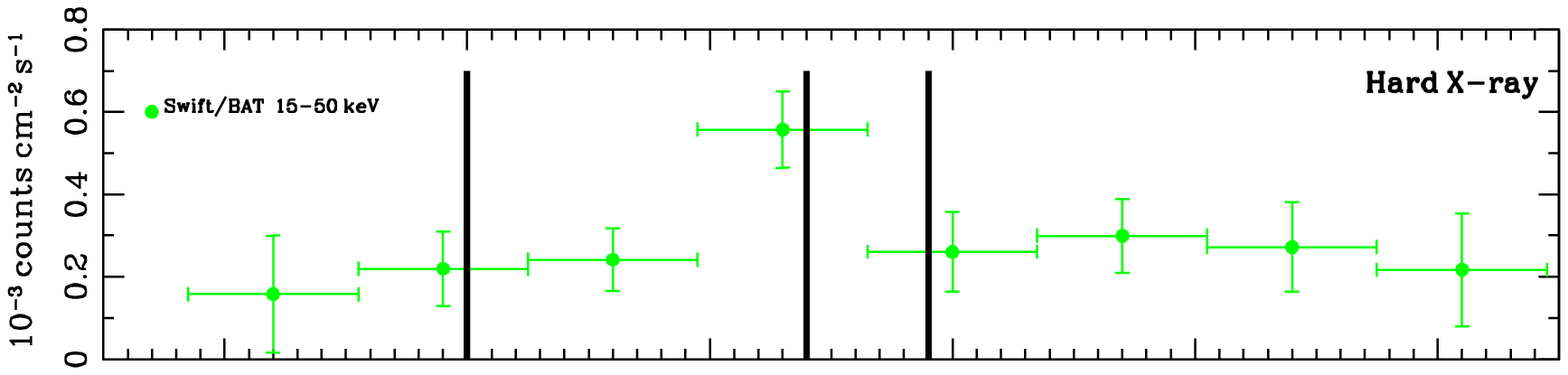} \\
\vspace{-15cm}
\includegraphics[width=7.5in]{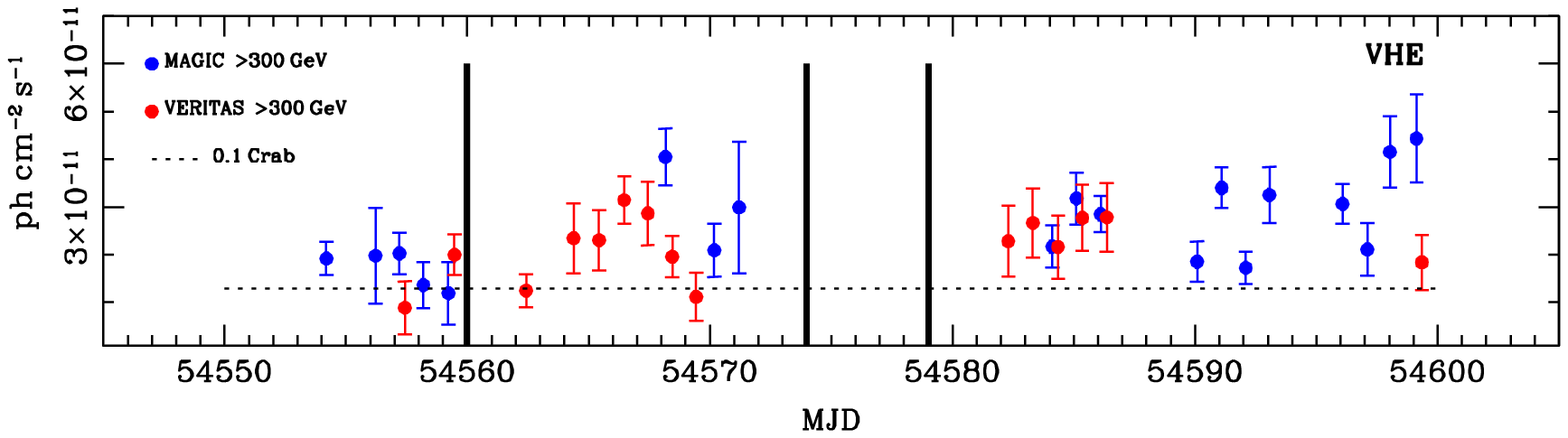}

\vspace{-14cm}
   \caption{Multifrequency light curve for Mrk\,501 during the entire
     campaign period. The panels from top to bottom show the radio,
     optical and UV, X-ray, hard X-ray, and VHE $\gamma$-ray bands. 
     The thick black vertical lines in all the panels
     delimit the time intervals corresponding to the three different
     epochs (P1, P2, and P3) used for the SED model fits in Sect. \ref{SEDSec}. The
     horizontal dashed line in the bottom panel depicts 10\% of the flux of the Crab
     nebula above 300 GeV \citep{Albert2008}.
   }
  \label{fig:lc}
\end{figure*}

Figure\,\ref{fig:lc} shows the light curves for all of the instruments
that participated in the campaign.  The five panels from top to bottom
present the light curves grouped into five energy ranges: radio,
optical, X-ray, hard X-ray, and VHE. 
%Given the slight differences in
%the frequencies covered by \Swiftc/UVOT and optical telescopes
%from  GASP-WEBT (R band), we scaled the later by 1/3 in order to
%display all these optical/UV flux measurements in the same panel.  

The multifrequency light curves show little variability; during this
campaign there were no outbursts of the magnitude observed in the past for
this object \citep[e.g.,][]{Krawczynski2000,Albert2007}.  
Around MJD 54560, there is an increase in the  X-rays activity, with 
a \Swiftc/XRT flux (in the energy range 0.3--10 keV) of
$\sim$1.3$\cdot$10$^{-10}$ erg cm$^{-2}$s$^{-1}$ before, and 
$\sim$1.7$\cdot$10$^{-10}$ erg cm$^{-2}$s$^{-1}$ after this day.
The measured X-ray flux during this campaign is well below 
$\sim$2.0$\cdot$10$^{-10}$ erg cm$^{-2}$s$^{-1}$, which is 
the average
X-ray flux measured with \Swiftc/XRT during the time interval of 2004
December 22 through 2012 August 31, which was reported in \citet{Stroh2013}.
In the VHE domain, the $\gamma$-ray flux above 300 GeV is mostly below 
$\sim$2$\cdot$10$^{-11}$ ph cm$^{-2}$s$^{-1}$ before MJD 54560, and above
$\sim$2$\cdot$10$^{-11}$ ph cm$^{-2}$s$^{-1}$ after this day.
The variability in the multifrequency
activity of the source is discussed in Sect. \ref{VarSec},
while the correlation among energy bands is reported in Sect.
\ref{CorrSec}.

For the spectral analysis presented in Sect. \ref{SEDSec}, we divided the
data set into three time intervals according to the X-ray flux
level (i.e., low/high flux before/after MJD 54560) 
and the data gap at most frequencies in the time interval MJD
54574-54579 (which is due to the difficulty of observing with IACTs 
during the nights with moonlight).
% (see Fig~\ref{fig:lc}).

\section{Variability}
\label{VarSec}

We followed the description given by \cite{Vaughan2003} to quantify
the flux variability by means of the fractional variability
parameter $F_{\mathrm{var}}$. To account for the individual
flux measurement errors ($\sigma_{\mathrm{err, i}}$), the `excess
variance' \citep{Edelson2002} was used as an estimator of
the intrinsic source flux variance. This is the variance after
subtracting the contribution expected from measurement statistical
uncertainties. This analysis does not account for systematic
uncertainties. $F_{\mathrm{var}}$ was derived for each participating
instrument individually, which covered an energy range from radio
frequencies at $\sim$8 GHz up to very high energies at $\sim$10
TeV. $F_{\mathrm{var}}$ is calculated as

\begin{equation}
  F_{\mathrm{var}} = \sqrt{\frac{S^2 - <\sigma_{\mathrm{err}}^2  >}{<F_{\gamma}>^2}} ~~~~~,
  \label{form_nva}
\end{equation}
where  $<F_{\gamma}>$ denotes the average photon flux, $S$ the
standard deviation of the $N$ flux measurements and
\mbox{$<\sigma_{\mathrm{err}}^2>$} the mean squared error, all
determined for a given instrument (energy bin). 
The uncertainty of
$F_{\mathrm{var}}$ is estimated according to

\begin{equation}
\Delta F_{\mathrm{var}} = \sqrt{F^{2}_{\mathrm{var}} +
  err(\sigma^{2}_{\mathrm{NXS}})} -F_{\mathrm{var}} ~~~~~~, \nonumber 
\end{equation}

where $err(\sigma^{2}_{\mathrm{NXS}})$ is given by equation 11 in \cite{Vaughan2003}, 
\begin{equation}
err(\sigma^{2}_{\mathrm{NXS}}) = \sqrt{\left(\sqrt{\frac{2}{N}}\frac{<\sigma^{2}_\mathrm{err}>}{<F_{\gamma}>^{2}}\right)^{2}  +
  \left(\sqrt{\frac{<\sigma^{2}_\mathrm{err}>}{N}}\frac{2F_{\mathrm{var}}}{<F_{\gamma}>}\right)^{2}} ~. \nonumber
\end{equation}

%\begin{equation}
%\Delta F_{\mathrm{var}} = \frac{\sqrt{<\sigma_{\mathrm{err}}^2>}}{\sqrt{N}
 % <F_{\gamma}>} \cdot \sqrt{1 + \frac{<\sigma_{\mathrm{err}}^2>}{2<F_{\gamma}>^2
 %   F_{\mathrm{var}}^2}} \nonumber
%\end{equation}

As reported in Sect. 2.2 in \cite{Poutanen2008}, this prescription
of computing $\Delta F_{\mathrm{var}}$ is more appropriate than that given by equation B2 in \cite{Vaughan2003}, which is not correct  when the error in the excess
variance is similar to or larger than the excess variance.
For this data set, we found that the prescription from
\cite{Poutanen2008}, which is used here, leads to $\Delta
F_{\mathrm{var}}$ that are $\sim$40\% smaller than those computed
with equation B2 in \cite{Vaughan2003} 
for the energy bands with the lowest $\frac{F_{\mathrm{var}}}{\Delta
  F_{\mathrm{var}}}$, while for most of the data points (energy bands)
the errors  are only $\sim$20\% smaller, and for the data points with the highest $\frac{F_{\mathrm{var}}}{\Delta
  F_{\mathrm{var}}}$ they are only few \% smaller.

\begin{figure}[]
  \centering
  \includegraphics[width=0.5\textwidth]{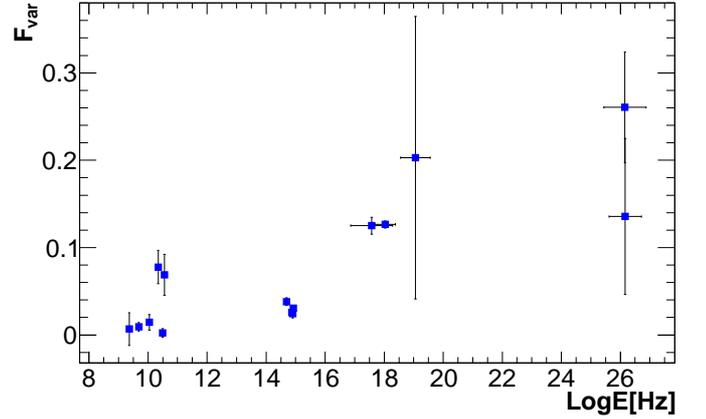}
  \caption{Fractional variability parameter $F_{\mathrm{var}}$ vs
    energy covered by the various instruments. $F_{\mathrm{var}}$
    was derived using the individual single-night flux measurements
    except for \Swiftc/BAT, for which, because of the limited sensitivity,
    we used data integrated over one week. Vertical bars denote $1\ \sigma$ uncertainties, horizontal
    bars indicate the approximate energy range covered by the
    instruments. }
  \label{fig_fracvar}
\end{figure}

Fig. \ref{fig_fracvar} shows the $F_{\mathrm{var}}$ values derived for
all instruments that participated in the MW campaign. 
The flux values that were used are displayed in
Fig.~\ref{fig:lc}. All flux values correspond to measurements
performed on minutes or hour timescales, except for \Swiftc/BAT, whose
X-ray fluxes correspond to a seven-day integration because of the somewhat moderate
sensitivity of this instrument to detect Mrk\,501. Consequently, 
\Swiftc/BAT data cannot probe the variability on timescales as short
as the other instruments, and hence $F_{\mathrm{var}}$  might be
underestimated for this instrument.  We obtained negative excess
variance ($<\sigma_{\mathrm{err}}^2>$ larger than $S^2$) for the
lowest frequencies of several radio telescopes. A negative excess
variance can occur when
there is little variability (in comparison with the uncertainty of the
flux measurements) and/or when the errors are slightly
overestimated. A negative excess variance can be interpreted as no
signature for variability in the data of that particular instrument,
either because a) there was no variability or b) the instrument was
not sensitive enough to detect it.  Fig. \ref{fig_fracvar}  only shows
the fractional variance for instruments with positive excess
variance. 

At radio frequencies, there is essentially no variability: all bands and
instruments show $F_{\mathrm{var}}$ close to zero, with the
exception of the of RATAN (22 GHz) and Mets\"ahovi (37 GHz), which show
$F_{\mathrm{var}} \sim ~7\pm2\%$. A possible reason for this
\emph{apparently} significant variability is unaccounted-for
errors due to variable weather conditions, which can easily add a
random extra fluctuation (day-by-day) of a few percent. 
However, it is worth mentioning that this flickering behavior has
been observed several times with Mets\"ahovi 
at 37 GHz, for example, in Mrk\,501 and also in Mrk\,421, while it is rare in other types of blazar objects; hence there is a chance that the measured fractional variability is dominated by a real flickering in the high-frequency radio emission of Mrk\,501. More studies on this aspect will be reported elsewhere.

During the 2008 campaign on Mrk\,501, we measured variability in the
optical, X-ray, and gamma-ray energy bands. The plot also shows some evidence
that the observed flux variability increases with energy: in the
optical $R$ band (ground-based telescopes) and the three UV filters from
\Swiftc/UVOT the variability is  $\sim$3\%, at X-rays it is $\sim$
13\%, and at VHE it is $\sim$20\%, although affected by relatively large error
bars (because of the statistical uncertainties in the individual flux measurements).

\section{Multifrequency cross-correlations \label{sec:cross} }
\label{CorrSec}

\begin{figure}[!t]
  \centering
  \subfloat[\RXTEc/PCA  vs. \Swiftc/XRT]{\includegraphics[width=2.6in]{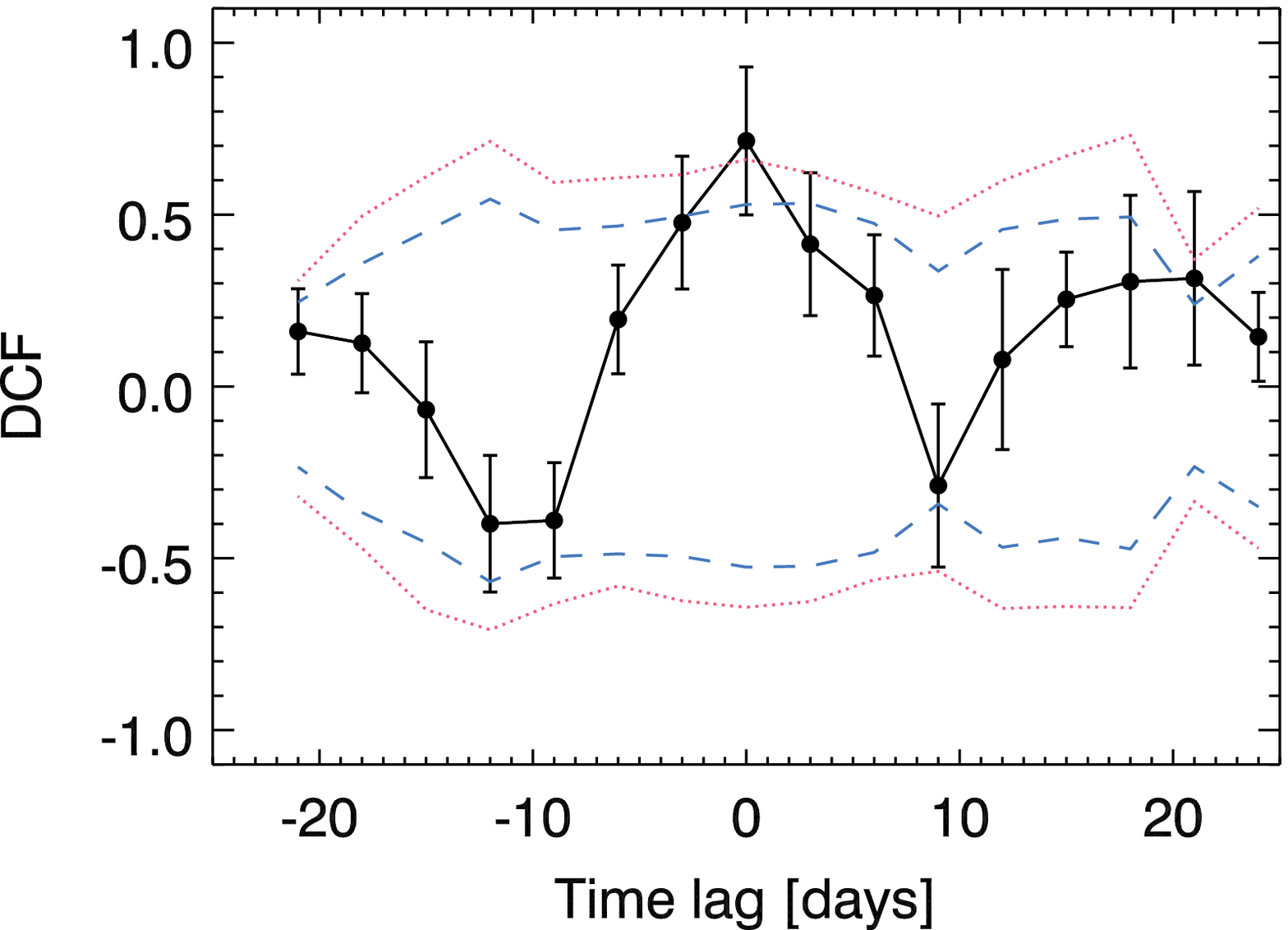}
    \label{fig_dcf1}} \hfil
  \subfloat[\RXTEc/PCA vs. MAGIC \& VERITAS]{\includegraphics[width=2.6in]{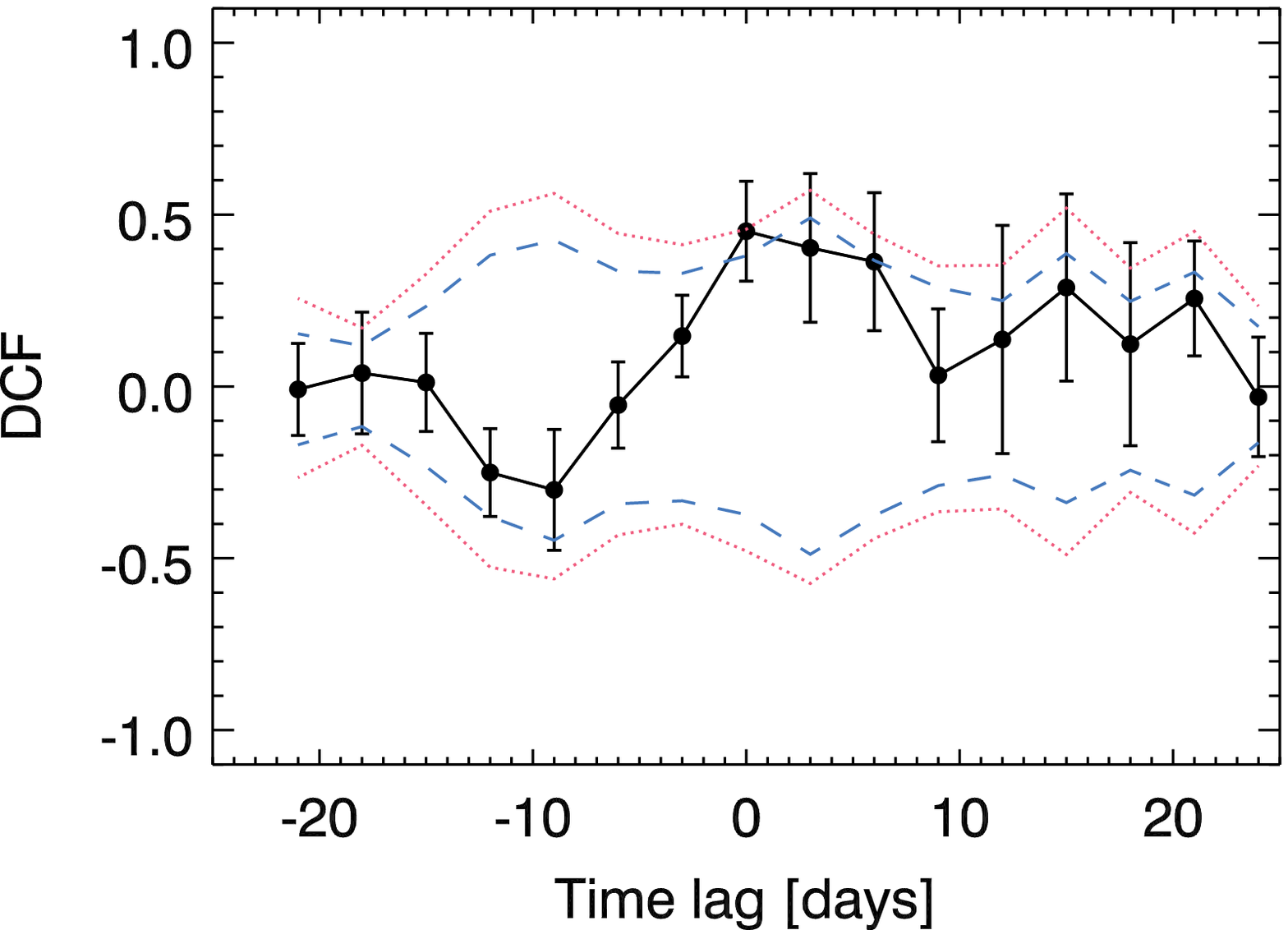}
    \label{fig_dcf2}}
  \caption{Discrete correlation function for time lags from -21 to +21 
    days in steps of 3 days. The (black) data points and errors are the DCF
    values computed according to the prescription given by
    \cite{Edelson1988}. The (blue) dashed and the (red) dotted curves
    depict the 95\% and 99\% confidence intervals for random correlations resulting from the
    dedicated Monte Carlo analysis described in Sect. \ref{CorrSec}.
}
\end{figure}

We used the discrete correlation function (DCF)
proposed by \cite{Edelson1988} to study the multifrequency cross-correlations between the
different energy bands. The DCF quantifies the 
temporal correlation as a function of the time lag between two light curves,
which can give us a deeper insight into the acceleration processes in the source.
For example, these time lags may occur as a result of spatially separated 
emission regions of the individual flux components (as expected, for example, 
in external inverse Compton models), or may be caused by the energy-dependent
cooling time-scales of the emitting electrons.

There are two important properties of the DCF method. First, it can be applied
to unevenly sampled data (as in this campaign), meaning that the correlation function
is defined only for lags for which the measured data exist, which
makes an interpolation of the data unnecessary. 
The result is a correlation function that is a set of discrete points binned in time.
Second, the errors in the individual flux measurements (which
contribute to the dispersion in the flux values) are naturally taken into
account. The latter characteristic is a big advantage over
the commonly used Pearson correlation function.
The main caveat of the DFC method is that the correlation function is
not continuous and that care needs to be taken when defining the time bins to achieve a reasonable balance between the required time resolution
and accuracy of DCF. Given the many two-day (sometimes three-day) time
gaps in the X-ray and VHE observations from this MW
campaign (see Figs.~\ref{fig:TimeEnergyCoverage} and \ref{fig:lc}), we
selected a time bin of three days to compute the DCF with minimal
impact of these observational gaps. Moreover, given the relatively low variability
reported in Fig.~\ref{fig:lc}, an estimation of DCF would not benefit
from a smaller time bin.

Using the data collected in this campaign, we derived the DCF for all different combinations of instruments and
energy regions and also for artificially introduced time lags (ranging
from -21 to +21 days) between the individual light curves.
Significant correlations were found only for the pairs \RXTEc/PCA - \Swiftc/XRT 
and also (less significant) \RXTEc/PCA with MAGIC and VERITAS
(Figs. \ref{fig_dcf1} and \ref{fig_dcf2}). In both cases, the highest
DCF values are obtained for a zero time lag, with a value of $0.71\pm0.22$
(\RXTEc/PCA - \Swiftc/XRT ) and $0.45\pm0.15$ (\RXTEc/PCA - MAGIC and
VERITAS), which implies positive correlations with a significance of
3.2 and 3.0 standard deviations.

As discussed in \cite{Uttley2003}, the errors in the DCF computed as
prescribed in
\cite{Edelson1988} might not be appropriate for determining the
significance of the DCF when the individual light-curve data points are
correlated red-noise data. Depending on the power spectral density (PSD) and the sampling
pattern, the significance as calculated by \cite{Edelson1988} might
therefore overestimate the real significance. 
To derive an independent estimate of the real significance of
the correlation peaks we used the dedicated Monte Carlo approach described below.

First we generated a large set of simulated light curves using the
method of \cite{Timmer1995} following the prescription of
\cite{Uttley2002}. As a model for the PSD we
assumed a simple power-law shape\footnote{
The shape of the PSD from blazars can be typically characterized with
a power law $P_{\nu} \propto \nu^{-\alpha}$ with spectral indices $\alpha$ between 1 and 2
\citep[see][]{Abdo2010Variability, Chatterjee2012}.}, and generated for each observed
light curve and for each PSD model (we varied the PSD slope in the
range -1.0 to -2.5 in steps of 0.1) 1000 simulated light curves. The
simulated light curves were then resampled using the sampling pattern
of the observed light curve. By applying the {\it psresp} method
\citep{Uttley2002} we tried to determine the best-fitting model for the
PSD. This involves the following steps in addition to the light-curve
simulation and resampling: the PSD of the observed light curve, as well
as the PSD of each simulated light curve, is calculated as the
square of the modulus of the discrete Fourier transform of the (mean
subtracted) light curve, as prescribed in \citet{Uttley2002}.
A $\chi^2$ analysis is
then used to determine the model that best fits the data. Given the
short frequency range, the uneven sampling and the presence of large
gaps (particularly in the VHE data), it was not possible to constrain
the PSD shape very tightly. The best-fitting models are power laws
with indices 1.4 (VHE) and 1.5 (X-rays), however, any power law with
an index between 1.0 and 1.9 fits the data reasonably well. The
\RXTEc/PCA light curve is sampled more often and regularly than the other
VHE and X-ray light curves, and moreover, \cite{Kataoka2001} found an X-ray PSD
slope similar to ours (1.37 $\pm$ 0.16) in the frequency range probed
here. Therefore we used the simulated \RXTEc/PCA light curves with a PSD slope
of -1.5 to ascertain the confidence levels in the DCF calculation. We
cross-correlated each of the 1000 simulated \RXTEc/PCA  light curves with
the observed VHE (MAGIC\&VERITAS) and \Swiftc/XRT light curves. The 95
and 99\% limits of the distribution of simulated \RXTEc/PCA
light curves when correlated with the real VHE and \Swiftc/XRT
light curves are plotted in Figs. \ref{fig_dcf1} and \ref{fig_dcf2} as blue dashed and red
dotted lines, respectively. The correlation peaks at time lag = 0 are 
higher than $>$ 99\% of the simulated data for the DCF for \RXTEc/PCA
correlated with \Swiftc/XRT, and $\sim$99\% for the simulated data for
the DCF for \RXTEc/PCA with VHE (MAGIC\&VERITAS). Given that a 99\%
confidence level is equivalent to 2.5 standard deviations, this result agrees reasonably well with the
significances of $\sim$3 standard deviations estimated from the Edelson\&Krolik DCF
errors, thus indicating that in this case the red-noise nature and the sampling of the light curve do not have a very strong influence. There are no other peaks or dips in the DCF between VHE and X-rays that appear significant.

%Due to the modest flux variability and / or large flux errors, no strong conclusions can be drawn from this analysis.
The positive correlation in the fluxes from \Swiftc/XRT and \RXTEc/PCA is
expected because of the proximity (and overlap) of the energy coverage of
these two instruments (see Table\,\ref{TableWithInstruments}), while the
 correlated behavior between \RXTEc/PCA and MAGIC/VERITAS suggests
that the X-ray and VHE emission are co-spatial and produced by
the same population of high-energy particles. The correlation between
the X-ray and VHE band has been reported many times in the past 
\citep[e.g.,][]{Krawczynski2000, Tavecchio2001, Gliozzi2006, Albert2007}, but only
when Mrk\,501 showed flaring VHE  activity with VHE fluxes higher than
the flux of the Crab nebula. An X-ray/VHE correlation when the
source shows a VHE flux below 0.5 Crab has never been shown until
now.

%%%%%%%%%%%% SED fig %%%%%%%%

\begin{figure}[h!]

  \centering

  %\begin{tabular}{ccc}

\vspace{-1cm}
    \includegraphics[width=0.52\textwidth]{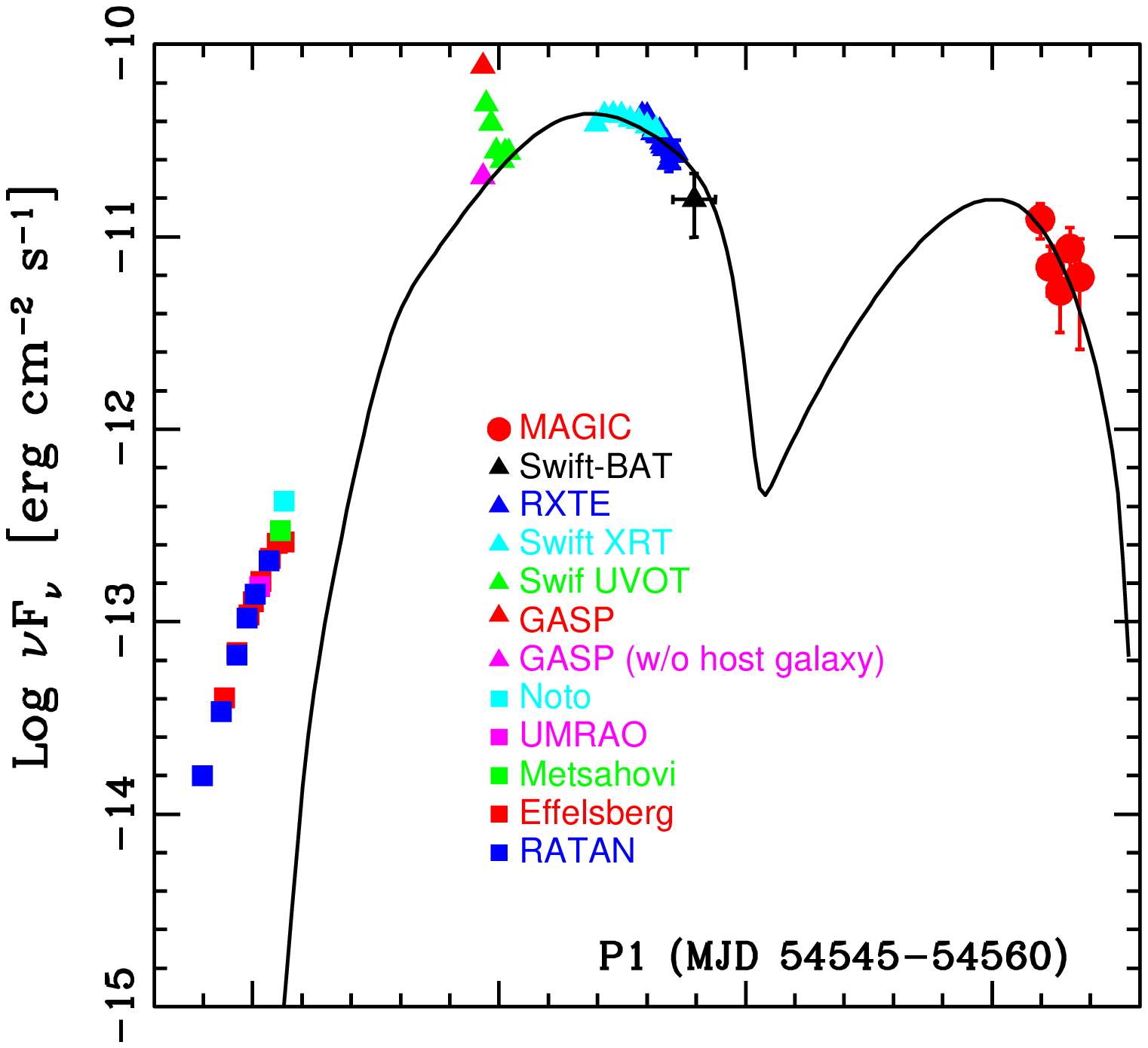}\\
\vspace{-2.9cm}

    \includegraphics[width=0.52\textwidth]{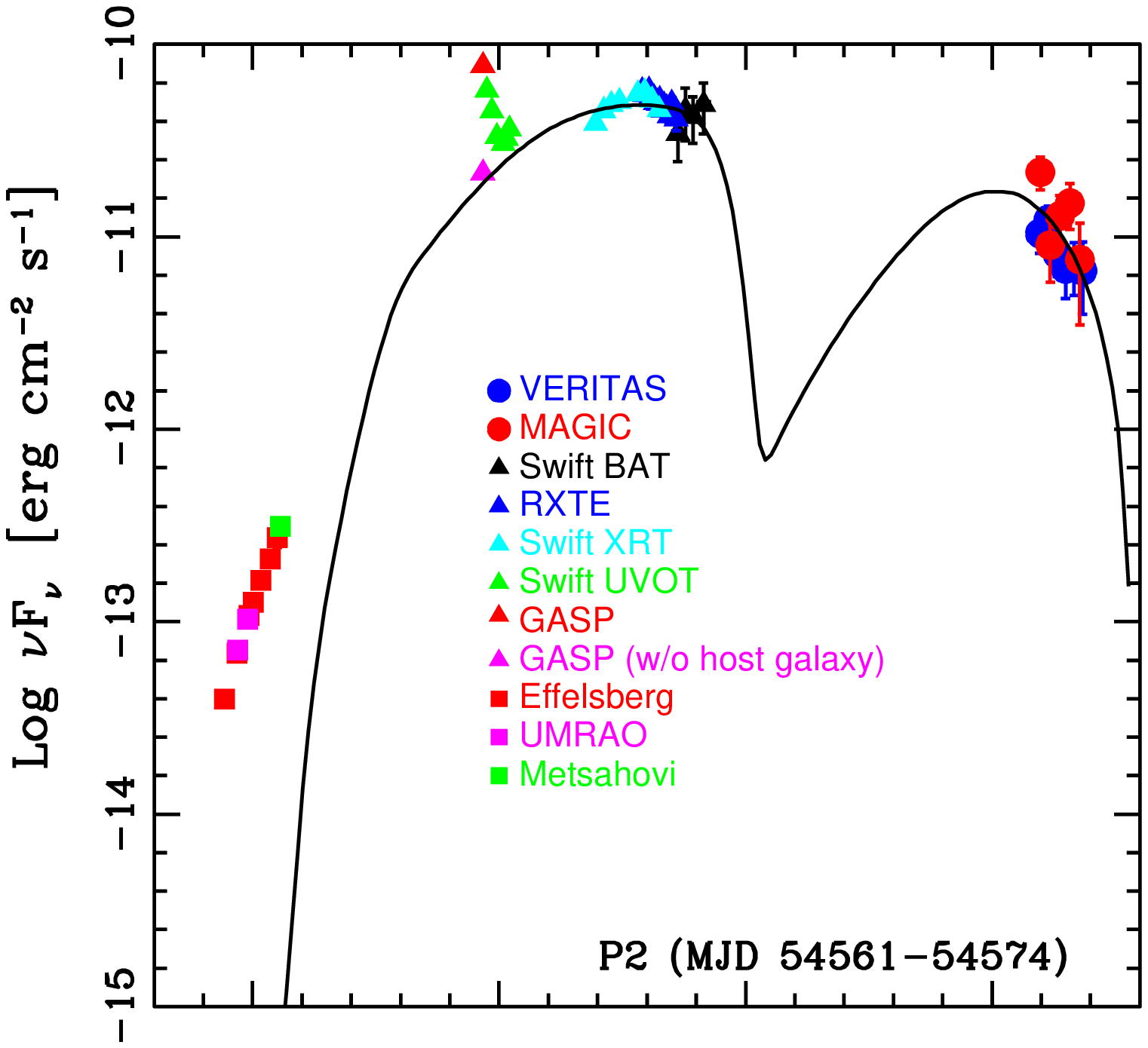}\\
\vspace{-2.9cm}

    \includegraphics[width=0.52\textwidth]{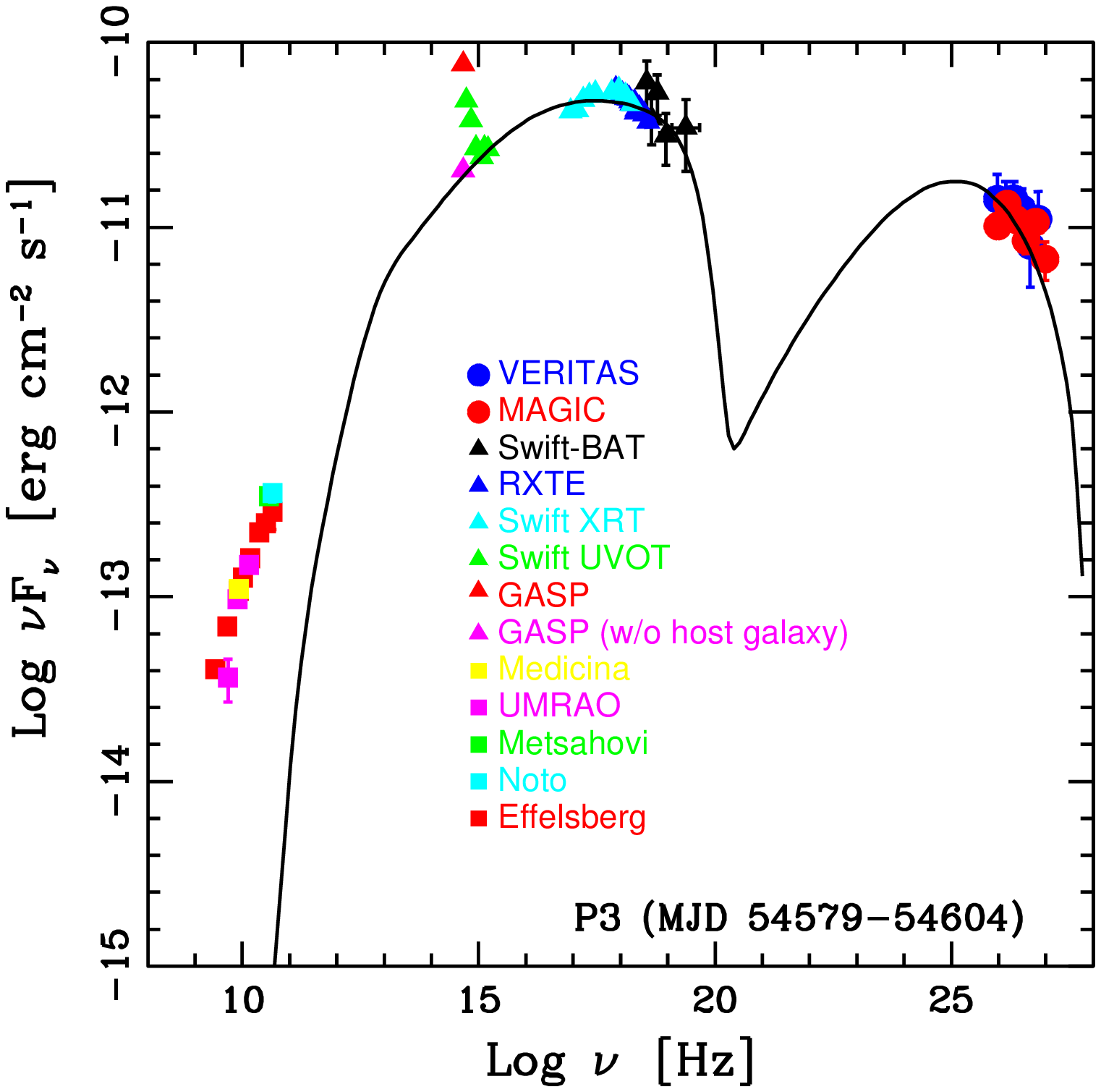}\\
    \vspace{-1cm}

  %\end{tabular}

  \label{fig:sed}
\caption{Spectral energy distributions for Mrk\,501 in the three
  periods described in Sect. \ref{LightCurve}. The legend reports the correspondence between the instruments
and the measured fluxes. Further details about the instruments are
given in Sect. \ref{sec_instr}. The vertical error bars in the data
points denote the 1 $\sigma$
statistical uncertainty.   The black curve depicts the one-zone
SSC model fit described in Sect. \ref{SEDSec}, with the resulting
parameters reported in Table\,\ref{sed_table}.}
\end{figure}

%%%%%%%%%%%%%%%%%%%%%%5

\section{SED modeling}
\label{SEDSec}
% http://heasarc.gsfc.nasa.gov/docs/heasarc/caldb/swift/docs/xrt/SWIFT-XRT-CALDB-09_v16.pdf

%RXTE (PCA)2% uncertainty -> Cui, W., et al., ApJL, 474, 57 (1997)
%Swift -> Cusumano, G, et al. Nuovo Cim., B121, 1463 (2006)
%MAGIC -> Aleksic et al., ApJ (in press)
 
Using the multifrequency data, we derived time-resolved SEDs for three
different periods that were defined according to the observed X-ray flux
during this campaign
(see Sect. \ref{LightCurve}).
The \Swiftc, \RXTEc, MAGIC, and VERITAS spectral results for the
  three periods are reported in Appendix \ref{XG_spectra}. The X-ray
  spectral results reported in Tables \ref{spectra_table_Swift} and \ref{spectra_table_RXTE} show that
  Mrk501 became brighter and harder in P2/P3 than in P1. 
  The VHE spectra reported in Tables \ref{spectra_table_MAGIC} and
  \ref{spectra_table_VERITAS} show that the MAGIC and VERITAS spectral
  results agree with each other within statistical uncertainties (despite
  the slightly different temporal coverage). The VHE spectral results do not show any significant spectral hardening when going
  from P1 to P2/P3. This could be due to the low VHE activity of
  Mrk501 and the moderate sensitivity that
  MAGIC and VERITAS had in 2008. In any case, MAGIC measures
  a VHE spectrum for P2/P3 that is significantly brighter than that measured for P1. 

The SED of the inner jet was modeled using a single-zone synchrotron self-Compton 
\citep[SSC,][]{Tavecchio1998, Maraschi2003} model, which is the simplest
theoretical framework for the broadband emission of
high-synchrotron-peaked BL Lac objects like Mrk\,501.
To reproduce the double bump 
shape of the SED, we assumed that the electron energy distribution
(EED) can be described by a broken 
power law, with indices $n_1$ and $n_2$, below and above the break ($\gamma_{\rm break}$), 
$\gamma_{\rm min}$ and $\gamma_{\rm max}$ being the lowest and highest
energies, and $K$ the normalization 
factor. The emission region is assumed to be a spherical plasmon of radius $R$, filled 
with a tangled homogeneous magnetic field of amplitude $B$, and moving with a relativistic Doppler factor $\delta$, 
such that $\delta=[\Gamma(1-\beta\, {\rm cos} \,\theta)]^{-1}$, where $\beta=v/c$,\, $\Gamma$ 
is the bulk Lorentz factor, and $\theta$ is the viewing angle with respect to the plasmon velocity.

The SED modeling was performed using a $\chi^{2}$ minimized fitting algorithm, instead of 
the commonly used $eye-ball$ procedure. The algorithm uses the
Levenberg-Marquardt method - which
interpolates between inverse Hessian method and steepest-descent method. 
In the fitting procedure, a systematic uncertainty of  $15\%$ for optical data sets,  $10\%$ for X-ray data sets, and  $40\%$ for VHE data sets was added in quadrature to the statistical uncertainty in the
differential energy fluxes.  The details of the fitting procedure can
be found in \cite{Mankuzhiyil2011}.
We note that the addition in quadrature
of the systematic and statistical errors to compute the overall
$\chi^{2}$  is not correct from a strictly statistical point of
view. Therefore, the $\chi^{2}$ was used as a penalty function for the
fit, and not as a measure of the true goodness-of-fit.
Consequently, even though the fitting algorithm allows us
to rapidly converge to a model that describes the data well,
the parameter errors provided by the fit are not statistically
meaningful, and hence were not used. 
%The fitting procedure is essentially used to get a faster (machine-driven)  convergence to a set of parameter values that describe well the overall broad-band emission of the source. 

The radio emission is produced by low-energy electrons, which can 
extend over hundreds of pc and even kpc distances, which is many
orders of magnitude larger than
the typical size of the regions where the blazar emission is produced
($\sim$10$^{-4}$--10$^{-1}$ pc). Given the relatively low angular
resolution of the single-dish radio telescopes (in comparison with
interferometric radio observations), these instruments measure the total
flux density of Mrk\,501 integrated over the whole source extension. 
Consequently, the single-dish radio data were used as upper
limits for the blazar emission modeled here.  The \Swiftc/UVOT
data points below 1.0$\times 10^{15}$\,Hz (those in the 
V, B, U filters) are dominated by the
emission from the host galaxy and hence they are
considered only as upper limits in the procedure of fitting 
the SED. The other \Swiftc/UVOT data points (those from the filters
UVW1, UVM2, and UVW2) were used in the SED model fit.
The optical data in the R band from GASP-WEBT were corrected for the
host galaxy contribution using the prescriptions from
\cite{Nilsson2007}, and the VHE data from MAGIC and VERITAS were corrected
for the absorption in the extragalactic background light (EBL) 
using the model from \cite{Franceschini2008}.  We note that, because of
the low redshift of this source, many other prescriptions 
\citep[e.g.,][]{Finke2010,Dominguez2011} 
provide compatible\footnote{At 5 TeV, most models predict an
  absorption of $\sim$0.4--0.5.} results at energies below 10 TeV.

%The data sets from all three periods were used in a first step in the
%minimization to evaluate the allowed parameter space. 
We noted that
the three SEDs can be described with minimal changes in the
environment parameters ($R$, $\delta$, $B$) and maximum energy of the
EED ($\gamma_{\rm max}$). Therefore, we decided to test whether we
could explain the modulations of the 
SED by simply changing the shape and normalization of the EED ($K$,
 $n1$, $n2$, $\gamma_{\rm break}$) while keeping all the other model parameters constant.
The collected multi-instrument data contain neither high-frequency ($>$43 GHz)
interferometric observations, nor \Fermic-LAT data and hence it is
difficult to constrain the model parameter  $\gamma_{\rm min}$. 
In fact, we noted that a one-zone SSC model can describe the
experimental data equally well with $\gamma_{\rm min}$=1 and 
$\gamma_{\rm min}$=1000. Both numbers have been used in the
literature, and the multi-instrument data from this campaign cannot be
used to distinguish between them. 
In this work we decided to use $\gamma_{\rm min}$=1000, which is motivated by
two reasons: {\it (i)} the preference for a large $\gamma_{\rm min}$
  in the one-zone SSC model fits in the Mrk\,501 SED
  reported in \cite{Abdo2011Mrk501}, where the experimental
constraints are tighter (because of usage of VLBA and \Fermi-LAT data); and {\it (ii)}
the preference for reducing the electron  energy density (which
largely depends on $\gamma_{\rm min}$ for 
soft-electron energy spectra) with respect to the magnetic energy density. 
We note that even with the choice of $\gamma_{\rm min}$=1000, the kinetic
(electron) energy density resulting from the SED model fit is about two orders of magnitude larger
than the magnetic energy density. 

The one-zone SSC model fits of the three different periods are shown in Fig. \ref{fig:sed}.  The resulting SED model parameters 
of the two scenarios are reported in Table \ref{sed_table}. The
relatively small variations in the broadband SED during this
observing campaign can be
adequately parameterized with small modifications in the parameters
that describe
the shape of the EED, namely $\gamma_{\rm
  break}$, $n_{1}$, $n_{2}$, and $K$. The one-zone SSC model parameters are
determined by the shape of the low-energy bump together with the overall
energy flux measured at VHE, and they are not sensitive to exact slope
of the VHE spectra. This is mostly due to the relatively large
uncertainties in the reported VHE spectra.

\begin{table*}
\begin{center}
  \begin{tabularx}{0.9\textwidth}{c c c c c c c c c c | c}
\hline\hline
\\
Period &  $\gamma_{\rm min}$ & $\gamma_{\rm break}$ &  $\gamma_{\rm max}$ &  $n_{1}$ & $n_{2}$ & ${B}$ [G] & $K$ [${\rm cm}^{-3}$] & ${R}$ [cm] & $\delta$ & Electron energy \\
& & & & & & & & & & density [erg ${\rm cm}^{-3}$]\\
\\
\hline
\\
%OLD
% P1 &  1.0$\times 10^{3}$ &  9.1$\times 10^{4}$  & 2.6$\times 10^{6}$ &  2.26  &  3.43  &  4.4$\times 10^{-2}$ &  2.7$\times 10^{4}$ &  1.1$\times 10^{16}$ &  21.5 & 1.0$\times 10^{-2}$\\
% P2 &  1.0$\times 10^{3}$ &  6.1$\times 10^{4}$  & 2.6$\times 10^{6}$ &  2.28  &  3.11  &  4.4$\times 10^{-2}$ &  3.8$\times 10^{4}$ &  1.1$\times 10^{16}$ &  21.5 & 1.2$\times 10^{-2}$\\
% P3 & 1.0$\times 10^{3}$ &  9.0$\times 10^{4}$  & 2.6$\times 10^{6}$ &  2.31  &  3.23  &  4.4$\times 10^{-2}$ &  4.5$\times 10^{4}$ &  1.1$\times 10^{16}$ &  21.5 & 1.2$\times 10^{-2}$\\

% NEW
P1 &  1.0$\times 10^{3}$ &  8.3$\times 10^{4}$  & 2.8$\times 10^{6}$ &  2.22  &  3.43  &  4.4$\times 10^{-2}$ &  2.1$\times 10^{4}$ &  9.7$\times 10^{15}$ &  22.8 & 1.1$\times 10^{-2}$\\
P2 &  1.0$\times 10^{3}$ &  4.6$\times 10^{4}$  & 2.8$\times 10^{6}$ &  2.23  &  3.09  &  4.4$\times 10^{-2}$ &  3.3$\times 10^{4}$ &  9.7$\times 10^{15}$ &  22.8 & 1.3$\times 10^{-2}$\\
P3 & 1.0$\times 10^{3}$ &  7.3$\times 10^{4}$  & 2.8$\times 10^{6}$ &  2.26  &  3.21  &  4.4$\times 10^{-2}$ &  3.6$\times 10^{4}$ &  9.7$\times 10^{15}$ &  22.8 & 1.3$\times 10^{-2}$\\

\\
\hline
  \end{tabularx}
  \caption{Model parameters obtained from the $\chi^2$-minimized SSC
    fits and the calculated electron energy density values.}
  \label{sed_table}
\end{center}
\end{table*}

\section{Discussion}
\label{discussion}

In the SSC framework, the observed flux
variability contains information on the dynamics of the underlying
population of relativistic electrons.
In this context, the general variability trend reported in
Fig. \ref{fig_fracvar} suggests that the flux variations are dominated
by the high-energy electrons, which have shorter cooling timescales,
which causes the higher variability amplitude
observed at the highest energies. 

Mrk~501 is known for its strong 
spectral variability at VHE; although these spectral
variations typically occur when the source's activity changes substantially,
showing a characteristic {\em harder-when-brighter} behavior
\citep[e.g.,][]{AharonianHEGRA2001,Albert2007,Abdo2011Mrk501}.
During this MW campaign the flux level and flux variability at VHE was low
(see Fig. \ref{fig:lc} and \ref{fig_fracvar}), and neither MAGIC nor VERITAS  
could detect significant spectral variability during the three temporal periods
considered (see Tables \ref{spectra_table_MAGIC} and
\ref{spectra_table_VERITAS}). This is partially due to the moderate
sensitivity of MAGIC and VERITAS back in 2008.
On the other hand, in the X-ray domain the instruments \Swiftc/XRT and
\RXTEc/PCA have sufficient sensitivity to resolve Mrk\,501 very significantly
in this very low state, and they both detect a hardening of
the spectra when the flux increases from P1 to P2 (see Tables \ref{spectra_table_Swift} and
\ref{spectra_table_RXTE}); this confirms the {\em harder-when-brighter} behavior reported previously for this source \citep[e.g.,][]{Gliozzi2006}.

The three SEDs  from the
2008 multi-instrument campaign can be adequately described
with a one-zone SSC model in which the EED
is parameterized with two power-law functions (i.e., one break).
Such a simple parameterization was not successful in describing the
SED from the 2009 multi-instrument campaign, which required
an EED described with three power-law functions
\citep{Abdo2011Mrk501}. 
This difference is related to the reduced instrumental energy coverage of the 2008 observing campaign in
comparison to that of 2009. In particular, the SED reported in
\cite{Abdo2011Mrk501} benefitted from 43 GHz VLBA interferometric
and 230 GHz SMA observations, as well as from {\it Fermi}-LAT, which
helped substantially to characterize the high-energy (inverse Compton)
bump. Therefore, the SEDs shown here have fewer experimental
constraints than those shown in \cite{Abdo2011Mrk501}, and this
might facilitate the characterization with a simpler theoretical model.
In addition, the somewhat higher activity of 
Mrk\,501 during 2009 than in 2008 is also worth mentioning, which might also
contribute to this difference in the SED modeling results. 

The obtained $\gamma_{\rm break}$ is $\sim$10 smaller
than the $\gamma_{\rm break}$ expected from synchrotron
cooling, which suggests that this break is intrinsic to the
injection mechanism. We note that this $\gamma_{\rm break}$ is 
comparable (within a factor of two) to the first  $\gamma_{\rm
  break}$ used in \cite{Abdo2011Mrk501}, 
which was also related to the mechanisms responsible for accelerating
the particles\footnote{The second break in the EED used in
  \cite{Abdo2011Mrk501} was related to the synchrotron
  cooling of the electrons.}.

Using the one-zone SSC model curves
presented in Sect. \ref{SEDSec}, we calculated the observed luminosity $L_{obs} = \int_{\nu_{\rm min}}^{\nu_{\rm max}} \! \nu\,F(\nu)$ with 
$\nu_{\rm min}=10^{11.0}$ and $\nu_{\rm max}=10^{27.5}$ Hz, and
converted it into jet power in radiation, $L_{r}=L_{obs}/\delta^2$, as
prescribed in \citet{Celotti2008}.
The radiated jet power for the three epochs were $6.2\times 10^{41}$ ${\rm erg}\,{\rm
  s}^{-1}$, $7.5\times 10^{41}$  ${\rm erg}\,{\rm s}^{-1}$,
and $7.4\times 10^{41}$ ${\rm erg}\,{\rm s}^{-1}$ for the periods\,1,
2, and 3 respectively; that is, the radiated jet power 
increased from P\,1 to P\,2 and remained the same from P\,2 to P\,3.
Given the model parameter
values reported in Table \ref{sed_table},  the increase
in the luminosity of the source is driven by a growth of the
electron energy density. In particular, the change
from P\,1 to P\,2 may have been produced by an injection of more electrons. On the other hand,
although in P\,3 we postulate slightly higher values of $K$ and $\gamma_{\rm break}$
than in P\,2, the softening of the electron spectrum ($n_2$) nullifies
the effect, such that the electron energy density, and hence the
luminosity, remain constant.

It is worth mentioning that the low X-ray and VHE activity reported in
this paper is comparable to the one reported 
for the MW campaign from 1996 March \citep{Kataoka1999}.  In this case, however, we could
describe the measured SED using a one-zone SSC with only one break
(instead of two) in the EED, and with a better data-model agreement at
VHE. The MAGIC and VERITAS spectra, after
being corrected for the absorption in the EBL, can be parameterized
with a power-law function with index $\sim$2.3, which matches the power-law index
predicted by the SSC model well, that is, $\sim$2.3 at 300 GeV and $\sim$2.5 at 1
TeV. On the other hand, the VHE spectra determined with HEGRA data from 1996 March to 1996 August
(hence not strictly simultaneous to the 1996 March MW campaign) was
parameterized with a power-law function with index $2.5\pm0.4$ above
1.5 TeV, which poorly matched the value of $\sim$3.8 predicted by
the SSC model used in \citet{Kataoka1999}. \citet{Kataoka1999} also postulated (based on comparisons of the
low-activity measured in 1996 with the large flare from 1997) that the
variability in the SED of Mrk~501 could be driven by variations in the
number of high-energy electrons.  Based on the collected broadband SEDs
of Mrk\,501 from 1997 to 2009, which were characterized with a one-zone SSC
scenario, \citet{Mankuzhiyil2012} also suggested that the variability
observed in this source is strongly related to the variability in the
high-energy portion of the EED.

%The luminosity of Mrk\,501 during this observing campaign, calculated within the one-zone SSC scenario, is the lowest luminosity ever
%observed in Mrk\,501. \citet{Mankuzhiyil2012} collected broadband SEDs
%of Mrk501 from 1997 to 2009 and characterized them within a one-zone SSC
%scenario, as we do here. The luminosities resulting from the model
%fits were caluclated by integrating $\nu\,F(\nu)$ from 2.5 decades below
%  the synchrotron peak until 0.75 decades above the inverse Compton
%  peak, finding values in the range from $3\times 10^{44}$ to $3\times
%  10^{45}$ ${\rm erg}\,{\rm s}^{-1}$. A similar calculation of the
%  luminosity from the SEDs presented here yield $\sim 3\times
%  10^{43}$ ${\rm erg}\,{\rm s}^{-1}$, hence an
%  order of magnitude lower luminosity than that computed for all the available MW samples used
%  in \cite{Mankuzhiyil2012}.

\section{Conclusions}
\label{conclusions}

We reported the results from a coordinated multi-instrument observation
of the TeV BL Lac Mrk\,501 between March and May 2008. This MW
campaign was planned regardless of the activity of source to
perform an unbiassed (by the high-activity) characterization of the
broadband emission. 

Mrk\,501 was found to be in a relatively low state of activity
with a VHE $\gamma$-ray flux of about 20\% the Crab nebula
flux. Nevertheless, significant flux variations were measured in
several energy bands, and a trend of variability increasing with
energy was also observed. 
We found a positive correlation between the 
activity of the source in the X-ray and VHE $\gamma$-ray bands.
The significance of this correlation was estimated with two
independent methods: {\it (i)} the prescription given in
\cite{Edelson1988}, and {\it (ii)} a tailored Monte Carlo approach
based on \cite{Uttley2002}. In both cases we found a marginally
significant ($\sim$3$\sigma$) positive correlation with zero time
lag. 
A X-ray to VHE correlation for Mrk\,501 has been reported many times
in the past during flaring (high) X-ray/VHE activity \citep[e.g.,][]{Krawczynski2000, Tavecchio2001, Gliozzi2006, Albert2007}; but this is the first time
that this behavior is reported for such a low X-ray/VHE state. Therefore this result suggests
that the mechanisms dominating the X-ray/VHE emission during
non-flaring activity do not differ substantially from those
that are
responsible for the emission during flaring activity.

We also showed that a homogeneous one-zone synchrotron self-Compton
model can describe
the Mrk\,501 SEDs measured during the two slightly different emission
states observed during this campaign. The difference between the
low (P1) and the slightly higher (P2 and P3) emission states can
be adequately 
modeled by changing the shape of the electron energy
distribution. 
But given the small variations in the broad band SED, other
combination of SSC parameter changes may also be able to describe the observations.

%This is in agreement with previous studies of the
%broadband emission of this source during flaring and non-flaring activity.

\section* {ACKNOWLEDGMENTS}

We are grateful to the journal referee, who helped us to improve
  the quality of this manuscript.
We would like to thank the Instituto de Astrof\'{\i}sica de
Canarias for the excellent working conditions at the
Observatorio del Roque de los Muchachos in La Palma.
The support of the German BMBF and MPG, the Italian INFN, 
the Swiss National Fund SNF, and the Spanish MICINN is 
gratefully acknowledged. This work was also supported by the CPAN CSD2007-00042 and MultiDark
CSD2009-00064 projects of the Spanish Consolider-Ingenio 2010
programme, by grant DO02-353 of the Bulgarian NSF, by grant 127740 of 
the Academy of Finland, by Projekt 09/176 of the Croatian science Foundation, 
by the DFG Cluster of Excellence ``Origin and Structure of the 
Universe'', by the DFG Collaborative Research Centers SFB823/C4 and SFB876/C3,
and by the Polish MNiSzW grant 745/N-HESS-MAGIC/2010/0.

This research is supported by grants from the U.S. Department of  
Energy Office of Science, the U.S. National Science Foundation and the  
Smithsonian Institution, by NSERC in Canada, by Science Foundation  
Ireland (SFI 10/RFP/AST2748) and by STFC in the U.K. We acknowledge  
the excellent work of the technical support staff at the Fred Lawrence  
Whipple Observatory and at the collaborating institutions in the  
construction and operation of the instrument.

The St.~Petersburg University team acknowledges support from Russian RFBR
foundation, grant 12-02-00452.
The Abastumani Observatory team acknowledges financial  support by the
Georgian National Science Foundation through grant GNSF/ST07/180.
The Mets\"ahovi team acknowledges the support from the Academy of Finland
to our observing projects (numbers 212656, 210338, 121148, and others).
We acknowledge the use of public data from the {\it Swift} and {\it RXTE} data archive.

\begin{appendix}
\label{app}

\section{X-ray and $\gamma$-ray spectra}
\label{XG_spectra}

This section reports the spectral parameters resulting from the fit to
the X-ray and $\gamma$-ray spectra.

\begin{table*}
\begin{center}
  \begin{tabularx}{0.6\textwidth}{c c c c c}
\hline\hline
\\
Period &  $K$ & $\alpha$ &  $\beta$ &  $\chi^2/ndf$ \\
& [10$^{-2}$ cm$^{-2}$ s$^{-1}$ keV${^-1}$] & & & \\
\\
\hline
\\
P1 &  2.65$\pm$0.03		&	2.01$\pm$0.01	& 0.24$\pm$0.03 & 331/308	\\
P2 & 3.12$\pm$0.03		&	1.85$\pm$0.01	& 0.23$\pm$0.02 &	322/336 \\
P3 &  3.23$\pm$0.03		&	1.87$\pm$0.01	& 0.26$\pm$0.02 & 409/354	\\
\\
\hline
  \end{tabularx}
  \caption{Parameters resulting from the fit with a log-parabola
    $F(E)=K \cdot (E/keV)^{- \alpha - \beta \cdot log(E/keV)}$ to the
    \Swiftc/XRT  spectra.}
  \label{spectra_table_Swift}
\end{center}
\end{table*}

\begin{table*}
\begin{center}
  \begin{tabularx}{0.5\textwidth}{c c c c}
\hline\hline
\\
Period &  $K$ & $\alpha$ &  $\chi^2/ndf$ \\
& [10$^{-2}$ cm$^{-2}$ s$^{-1}$ keV${^-1}$] & & \\
\\
\hline
\\
P1 &  4.36$\pm$0.21		&	2.36$\pm$0.03	& 24/19 \\
P2 & 4.69$\pm$0.18		&	2.19$\pm$0.02	& 18/19	\\
P3 &  4.78$\pm$0.10	        &	2.23$\pm$0.01	& 24/19	\\
\\
\hline
  \end{tabularx}
  \caption{Parameters resulting from the fit with a power law
    $F(E)=K \cdot (E/keV)^{- \alpha}$ to the
    \RXTEc/PCA  spectra.}
  \label{spectra_table_RXTE}
\end{center}
\end{table*}

\begin{table*}
\begin{center}
  \begin{tabularx}{0.5\textwidth}{c c c c}
\hline\hline
\\
Period &  $K$ & $\alpha$ &  $\chi^2/ndf$ \\
& [10$^{-12}$ cm$^{-2}$ s$^{-1}$ TeV${^-1}$] & & \\
\\
\hline
\\
P1 &  5.3$\pm$0.5		&	2.49$\pm$0.20	& 5/3 \\
P2 &  9.1$\pm$0.8		&	2.44$\pm$0.17	& 5/3	\\
P3 &  7.7$\pm$0.3	        &	2.37$\pm$0.05	& 9/4	\\
All &  7.4$\pm$0.2	&	2.42$\pm$0.05	& 2/4	\\
\\
\hline
  \end{tabularx}
  \caption{Parameters resulting from the fit with a power law
    $F(E)=K \cdot (E/TeV)^{- \alpha}$ to the
    measured MAGIC  spectra (without correction for the EBL absorption)}
  \label{spectra_table_MAGIC}
\end{center}
\end{table*}

\begin{table*}
\begin{center}
  \begin{tabularx}{0.5\textwidth}{c c c c}
\hline\hline
\\
Period &  $K$ & $\alpha$ &  $\chi^2/ndf$ \\
& [10$^{-12}$ cm$^{-2}$ s$^{-1}$ TeV${^-1}$] & & \\
\\
\hline
\\
P1 &  ---		&	---	& --- \\
P2 &  6.0$\pm$0.9		&	2.55$\pm$0.22	& 2/4	\\
P3 &  8.7$\pm$1.5	        &	2.44$\pm$0.28	& 1/4	\\
All &  9.4$\pm$0.6	&	2.47$\pm$0.10	& 13/8	\\
\\
\hline
  \end{tabularx}
  \caption{Parameters resulting from the fit with a power law
    $F(E)=K \cdot (E/TeV)^{- \alpha}$ to the measured
    VERITAS  spectra (without correction for the EBL absorption)}
  \label{spectra_table_VERITAS}
\end{center}
\end{table*}

\end{appendix}

\end{document}